\documentclass{article}
\usepackage{amssymb}
\usepackage{amsfonts}
\usepackage{amsmath}

\input{tcilatex}

\begin{document}

\begin{center}
\bigskip

{\large LORENTZIAN CR STRUCTURES}

{\large \ }\bigskip

C. N. Ragiadakos

IEP, Tsoha 36, Athens 11521, Greece

email: crag@iep.edu.gr , ragiadak@gmail.com

\bigskip

\textbf{ABSTRACT}

The mathematics of a 4-dimensional renormalizable generally covariant
lagrangian model (with first order derivatives) is reviewed. The lorentzian
CR manifolds are totally real submanifolds of 4(complex)-dimensional complex
manifolds determined by four special conditions. The defining tetrad permits
the definition of a class of lorentzian metrics which admit two geodetic and
shear free congruences. These metrics permit the classification of the
structures using the Weyl tensor and the Flaherty pseudo-complex structure.
The Cartan procedure permits the definition of three relative invariants and
the osculation of the generic lorentzian CR structures with the U(2)
manifold. Two hypersurface-type 3-dimensional CR structures, related with a
parametric anti-holomorphic relation, consist a lorentzian CR structure. An
osculation on the basis of the SU(2,2) group reveals the Poincar\'{e} group,
which may be identified with the observed group in nature. This possibility
may differentiate the lorentzian CR manifolds into realizable particle-like
and non-realizable unparticles. Examples of static axially symmetric
lorentzian CR structures are computed. For every lorentzian CR manifold, a
class of Kaehler metrics in the ambient complex manifold is found, which
induce the class of compatible lorentzian metrics on the submanifold. Then
the lorentzian CR manifold becomes a lagrangian submanifold in the
corresponding ambient (Kaehler) symplectic manifold. In this context the
lorentzian CR manifolds may be considered as dynamical processes in the view
of the Einstein-Infeld-Hofman derivation of the equations of motion, which
may provide an interaction picture analogous to the splitting and merging of
2-dimensional surfaces in String Field Theory.
\end{center}

\pagebreak

\bigskip

\bigskip

\bigskip

{\LARGE Contents}

\textbf{1. INTRODUCTION}

\qquad 1.1 The Kerr-Newman lorentzian CR structure

\textbf{2. RELATIVE INVARIANTS OF LORENTZIAN CR STRUCTURES}

\textbf{3. WEYL CURVATURE CLASSIFICATION OF LORENTZIAN CR STRUCTURES}

\textbf{4. FLAHERTY'S PSEUDO-COMPLEX STRUCTURE}

\textbf{5. A PAIR OF 3-DIMENSIONAL CR STRUCTURES}

\textbf{6. SURFACES OF THE SU(2,2) CLASSICAL DOMAIN}

\qquad 6.1 Asymptotically flat lorentzian CR manifolds

\qquad 6.2 The Poincar\'{e} group

\qquad 6.3 Complex trajectory emergence from lorentzian CR manifolds

\textbf{7. EXAMPLES OF SYMMETRIC LORENTZIAN CR MANIFOLDS}

\textbf{8. ON GEOMETRIC QUANTIZATION OF LORENTZIAN CR MANIFOLDS}

\textbf{9. TOWARDS A LORENTZIAN CR QUANTUM FIELD THEORY}

\pagebreak

\renewcommand{\theequation}{\arabic{section}.\arabic{equation}}

\section{INTRODUCTION}

\setcounter{equation}{0}

Einstein's theory of relativity is based on the riemannian geometry,
assuming the metric tensor $g_{\mu \nu }$ as its fundamental dynamical
variable. But it is actually well known that the conventional quantization
of the Einstein-Hamilton action does not imply a self-consistent quantum
field theory of gravity. Many years ago, trying to transfer in four
dimensions the metric independence property of the linearized string action%
\cite{POL} 
\begin{equation}
I_{S}=\frac{1}{2}\int d^{2}\!\xi \ \sqrt{-\gamma }\ \gamma ^{\alpha \beta }\
\partial _{\alpha }X^{\mu }\partial _{\beta }X^{\nu }\eta _{\mu \nu }=\int
d^{2}\!z\ \partial _{0}X^{\mu }\partial _{\widetilde{0}}X^{\nu }\eta _{\mu
\nu }  \label{i1}
\end{equation}%
I found the following four dimensional special Yang-Mills action\cite%
{RAG2013} 
\begin{equation}
\begin{array}{l}
I_{G}=\frac{1}{2}\int d^{4}\!z\det (g_{\alpha \widetilde{\alpha }})\
g^{\alpha \widetilde{\beta }}g^{\gamma \widetilde{\delta }}F_{\!j\alpha
\gamma }F_{\!j\widetilde{\beta }\widetilde{\delta }}+c.\ conj.=\int
d^{4}\!z\ F_{\!j01}F_{\!j\widetilde{0}\widetilde{1}}+c.\ c. \\ 
\\ 
F_{j_{ab}}=\partial _{a}A_{jb}-\partial _{a}A_{jb}-\gamma
\,f_{jik}A_{ia}A_{kb}%
\end{array}
\label{i2}
\end{equation}%
which does not depend on the metric. I call this lagrangian model quantum
cosmodynamic. But there is an essential difference between the two and four
dimensional cases. In two dimensions any orientable surface admits a metric,
which takes the form $ds^{2}=2dz^{0}dz^{\widetilde{0}}$. In four dimensions
a metric takes the convenient form $ds^{2}=2g_{\alpha \widetilde{\beta }%
}dz^{\alpha }dz^{\widetilde{\beta }}$, if it admits two geodetic and shear
free null congruences. This restricts the spacetimes to a special kind of
totally real submanifolds\cite{BAOU} of a complex manifold, which I will
call lorentzian CR manifolds (LCR-manifold). This metric independence makes
the quantum cosmodynamic model formally renormalizable. I think that it is
the unique known generally covariant renormalizable model with first order
derivatives in four dimensions. In this paper I will review the mathematical
properties of these LCR-manifolds in order to make the problems of the
quantum cosmodynamic model more comprehensive to mathematicians. A review of
the physical picture of the model is given elsewhere\cite{RAG2013}.

In the context of the Newman-Penrose (NP) formalism in General relativity we
usually work with the tangent space basis (tetrad) ($\ell ^{\mu }\partial
_{\mu }\ ,\ n^{\mu }\partial _{\mu }\ ,\ m^{\mu }\partial _{\mu }\ ,\ 
\overline{m}^{\mu }\partial _{\mu }$), where the first two are real, the
third one is complex and the forth one is the complex conjugate of the third
vector, as the notation indicates. The following commutation relations
define the NP coefficients

\begin{equation}
\begin{array}{l}
\lbrack \ell ^{\mu }\partial _{\mu }\ ,\ n^{\nu }\partial _{\nu }]=-(\gamma +%
\overline{\gamma })\ell ^{\rho }\partial _{\rho }-(\varepsilon +\overline{%
\varepsilon })n^{\rho }\partial _{\rho }+(\overline{\tau }+\pi )m^{\rho
}\partial _{\rho }+(\tau +\overline{\pi })\overline{m}^{\rho }\partial
_{\rho } \\ 
\\ 
\lbrack \ell ^{\mu }\partial _{\mu }\ ,\ m^{\nu }\partial _{\nu }]=(%
\overline{\pi }-\overline{\alpha }-\beta )\ell ^{\rho }\partial _{\rho
}-\kappa n^{\rho }\partial _{\rho }+(\overline{\rho }+\varepsilon -\overline{%
\varepsilon })m^{\rho }\partial _{\rho }+\sigma \overline{m}^{\rho }\partial
_{\rho } \\ 
\\ 
\lbrack n^{\mu }\partial _{\mu }\ ,\ m^{\nu }\partial _{\nu }]=\overline{\nu 
}\ell ^{\rho }\partial _{\rho }+(\overline{\alpha }+\beta -\tau )n^{\rho
}\partial _{\rho }+(\gamma -\overline{\gamma }-\mu )m^{\rho }\partial _{\rho
}-\overline{\lambda }\overline{m}^{\rho }\partial _{\rho } \\ 
\\ 
\lbrack m^{\mu }\partial _{\mu }\ ,\ \overline{m}^{\nu }\partial _{\nu
}]=(\mu -\overline{\mu })\ell ^{\rho }\partial _{\rho }+(\rho -\overline{%
\rho })n^{\rho }\partial _{\rho }+(\overline{\beta }-\alpha )m^{\rho
}\partial _{\rho }+(\overline{\alpha }-\beta )\overline{m}^{\rho }\partial
_{\rho } \\ 
\end{array}
\label{i3}
\end{equation}%
The corresponding basis of the cotangent space ($\ell \ ,\ n\ ,\ m\ ,\ 
\overline{m}$) is defined via the relations

\begin{equation}
\begin{array}{l}
\ell ^{\mu }n_{\mu }=1\ ,\ m^{\nu }\overline{m}_{\nu }=-1\ ,\ all\ other\
contractions\ vanish \\ 
\end{array}
\label{i4}
\end{equation}%
I point out that I have not yet introduced any metric. The CR structure does
not need the notion of the metric. The conditions (\ref{i4}) are implied by
properly inverting the $4\times 4$ matrix ($\ell _{\mu }\ ,\ n_{\mu }\ ,\
m_{\mu }\ ,\ \overline{m}_{\mu }$). Then one can find the following
differential forms%
\begin{equation}
\begin{array}{l}
d\ell =-(\varepsilon +\overline{\varepsilon })\ell \wedge n+(\alpha +%
\overline{\beta }-\overline{\tau })\ell \wedge m+(\overline{\alpha }+\beta
-\tau )\ell \wedge \overline{m}- \\ 
\qquad -\overline{\kappa }n\wedge m-\kappa n\wedge \overline{m}+(\rho -%
\overline{\rho })m\wedge \overline{m} \\ 
\\ 
dn=-(\gamma +\overline{\gamma })\ell \wedge n+\nu \ell \wedge m+\overline{%
\nu }\ell \wedge \overline{m}+(\pi -\alpha -\overline{\beta })n\wedge m+ \\ 
\qquad +(\overline{\pi }-\overline{\alpha }-\beta )n\wedge \overline{m}+(\mu
-\overline{\mu })m\wedge \overline{m} \\ 
\\ 
dm=-(\tau +\overline{\pi })\ell \wedge n+(\gamma -\overline{\gamma }+%
\overline{\mu })\ell \wedge m+\overline{\lambda }\ell \wedge \overline{m}+
\\ 
\qquad \quad +(\varepsilon -\overline{\varepsilon }-\rho )n\wedge m-\sigma
n\wedge \overline{m}+(\beta -\overline{\alpha })m\wedge \overline{m} \\ 
\end{array}
\label{i5}
\end{equation}

\textbf{DEFINITION:} A regular tetrad ($\det (e_{\mu }^{a})\neq 0,\ \infty $%
) satisfying the relations $\kappa =\sigma =\lambda =\nu =0$ defines a
lorentzian CR structure (LCR-structure). Then we will say that the
corresponding manifold admits a LCR-structure and it will be called
lorentzian CR manifold (LCR-manifold).

Using the tetrad, the LCR-structure conditions take the form

\begin{equation}
\begin{array}{l}
(\ell ^{\mu }m^{\nu }-\ell ^{\nu }m^{\mu })(\partial _{\mu }\ell _{\nu
})=0\;\;\;\;,\;\;\;\;(\ell ^{\mu }m^{\nu }-\ell ^{\nu }m^{\mu })(\partial
_{\mu }m_{\nu })=0 \\ 
\\ 
(n^{\mu }m^{\nu }-n^{\nu }m^{\mu })(\partial _{\mu }n_{\nu
})=0\;\;\;\;,\;\;\;\;(n^{\mu }m^{\nu }-n^{\nu }m^{\mu })(\partial _{\mu
}m_{\nu })=0%
\end{array}
\label{i6}
\end{equation}%
Then Frobenius theorem states that there are four independent complex
functions $(z^{\alpha },\;z^{\widetilde{\alpha }})$,\ $\alpha =0,\ 1$ , such
that

\begin{equation}
\begin{array}{l}
dz^{\alpha }=f_{\alpha }\ \ell _{\mu }dx^{\mu }+h_{\alpha }\ m_{\mu }dx^{\mu
}\;\;\;\;,\;\;\;dz^{\widetilde{\alpha }}=f_{\widetilde{\alpha }}\ n_{\mu
}dx^{\mu }+h_{\widetilde{\alpha }}\ \overline{m}_{\mu }dx^{\mu } \\ 
\\ 
\ell =\ell _{\alpha }dz^{\alpha }\;\;\;\;,\;\;\;m=m_{\alpha }dz^{\alpha } \\ 
\\ 
n=n_{\widetilde{\alpha }}dz^{\widetilde{\alpha }}\;\;\;\;,\;\;\;\overline{m}%
=m_{\widetilde{\alpha }}dz^{\widetilde{\alpha }} \\ 
\end{array}
\label{i7}
\end{equation}%
This LCR-structure is called realizable or embedable. A real analytic
LCR-manifold is always realizable. I do not actually know whether there are
non-realizable LCR-manifolds.

The reality conditions of the tetrad imply that the structure coordinates $%
z^{b}=(z^{\alpha },\;z^{\widetilde{\alpha }})$,\ $\alpha =0,\ 1$ satisfy the
relations

\begin{equation}
\begin{array}{l}
dz^{0}\wedge dz^{1}\wedge d\overline{z^{0}}\wedge d\overline{z^{1}}=0 \\ 
\\ 
dz^{\widetilde{0}}\wedge dz^{\widetilde{1}}\wedge d\overline{z^{0}}\wedge d%
\overline{z^{1}}=0 \\ 
\\ 
dz^{\widetilde{0}}\wedge dz^{\widetilde{1}}\wedge d\overline{z^{\widetilde{0}%
}}\wedge d\overline{z^{\widetilde{1}}}=0 \\ 
\\ 
dz^{0}\wedge dz^{1}\wedge dz^{\widetilde{0}}\wedge dz^{\widetilde{1}}\neq 0%
\end{array}
\label{i8}
\end{equation}%
that is, there are two real functions $\rho _{11}$ , $\rho _{22}$ and a
complex one $\rho _{12}$, such that

\begin{equation}
\begin{array}{l}
\rho _{11}(\overline{z^{\alpha }},z^{\alpha })=0\quad ,\quad \rho
_{12}\left( \overline{z^{\alpha }},z^{\widetilde{\alpha }}\right) =0\quad
,\quad \rho _{22}\left( \overline{z^{\widetilde{\alpha }}},z^{\widetilde{%
\alpha }}\right) =0 \\ 
\\ 
\frac{\partial \rho _{ij}}{\partial z^{b}}\neq 0\neq \frac{\partial \rho
_{ij}}{\partial \overline{z^{b}}} \\ 
\end{array}
\label{i9}
\end{equation}%
According to the conventional terminology, the manifold is locally a totally
real submanifold of $C^{4}$. These relations and the complex structure of
the ambient manifold determine a LCR-structure on the 4-dimensional
manifold. Notice that the defining functions do not depend on all the
structure coordinates. The precise dependence of the defining functions on
the structure coordinates characterizes the LCR-structure from the general
definition of a totally real submanifold of $C^{4}$.

The four functions $z^{b}\equiv (z^{\alpha },\;z^{\widetilde{\alpha }})$,\ $%
\alpha =0,\ 1$ are the structure coordinates of the (integrable) complex
structure. The holomorphic transformations which preserve the LCR-structure
are

\begin{equation}
\begin{array}{l}
z^{\prime \alpha }=f^{\alpha }(z^{\beta })\quad ,\quad z^{\prime \widetilde{%
\alpha }}=f^{\widetilde{\alpha }}(z^{\widetilde{\beta }}) \\ 
\end{array}
\label{i10}
\end{equation}%
I point out that the general holomorphic transformations $z^{\prime
b}=f^{b}(z^{c})$ do not preserve the LCR-structure!

The inverse procedure to find a tetrad ($\ell \ ,\ n\ ,\ m\ ,\ \overline{m}$%
) from the defining conditions (\ref{i9}) is straightforward. It is
convenient to use the notation $\partial ^{\prime }f=\frac{\partial f}{%
\partial z^{\alpha }}dz^{\alpha }$\ and $\partial ^{\prime \prime }f=\frac{%
\partial f}{\partial z^{\widetilde{\alpha }}}dz^{\widetilde{\alpha }}$.
Because of $d\rho _{ij}=0$ and the special dependence of each function on
the structure coordinates $\left( z^{\alpha },z^{\widetilde{\alpha }}\right) 
$, we find\cite{JACO} \ 
\begin{equation}
\begin{array}{l}
\ell =2i\partial \rho _{11}=2i\partial ^{\prime }\rho _{11}=i(\partial
^{\prime }-\overline{\partial ^{\prime }})\rho _{11}=-2i\overline{\partial }%
\rho _{11} \\ 
\\ 
n=2i\partial \rho _{22}=2i\partial ^{\prime \prime }\rho _{22}=i(\partial
^{\prime \prime }-\overline{\partial ^{\prime \prime }})\rho _{22}=-2i%
\overline{\partial ^{\prime \prime }}\rho _{22} \\ 
\\ 
m_{1}=2i\partial \frac{\rho _{12}+\overline{\rho _{12}}}{2}=i(\partial
^{\prime }+\partial ^{\prime \prime }-\overline{\partial ^{\prime }}-%
\overline{\partial ^{\prime \prime }})\frac{\rho _{12}+\overline{\rho _{12}}%
}{2} \\ 
\\ 
m_{2}=2i\partial \frac{\overline{\rho _{12}}-\rho _{12}}{2i}=i(\partial
^{\prime }+\partial ^{\prime \prime }-\overline{\partial ^{\prime }}-%
\overline{\partial ^{\prime \prime }})\frac{\overline{\rho _{12}}-\rho _{12}%
}{2i} \\ 
\end{array}
\label{i11}
\end{equation}%
where we consider all these forms restricted on the defined submanifold,
therefore they are real. The relations become simpler, if we use the complex
form \ 
\begin{equation}
\begin{array}{l}
m=m_{1}+im_{2}=2i\partial ^{\prime }\overline{\rho _{12}}=-2i\overline{%
\partial ^{\prime \prime }}\overline{\rho _{12}}=i(\partial ^{\prime }-%
\overline{\partial ^{\prime \prime }})\overline{\rho _{12}} \\ 
\end{array}
\label{i12}
\end{equation}

In the next subsection, the typical example of the Kerr-Newman LCR-structure
will be presented. In section II, the first three relative invariants of the
LCR-structure will be defined. In the case of non-vanishing relative
invariants, the LCR-structure is determined by two real and one complex
vector fields. It will also become clear that the static spherically
symmetric spacetimes have trivial LCR-structures. In sections III and IV, we
present classifications of the LCR-structures based on a regular compatible
metric and the Flaherty pseudo-complex structure. In section V we show that
a parameter dependent antiholomorphic transformation of a 3-dimensional CR
structure defines a LCR-structure. In section VI an osculation of the
LCR-structure is written down using the homogeneous coordinates of the $%
G_{2,2}$ grassmannian manifold. This reveals the Poincar\'{e} group. In
section VII, we use these Poincar\'{e} transformations to find static
axially symmetric LCR-manifolds. In section VIII, I find a class of Kaehler
metrics in the ambient complex manifold, which induce compatible metrics on
the submanifold. The LCR-submanifold is a lagrangian submanifold relative to
the corresponding Kaehler 2-form. Using the properties of
path(functional)-integral quantization of the LCR-structure dependent
action, we try, in section IX, to formulate an interacting quantum theory of
LCR-manifolds, analogous to the interacting String Field Theory of
2-dimensional surfaces.

\subsection{The Kerr-Newman LCR-structure}

The spacetimes with two geodetic and shear free null congruences are typical
examples of LCR-manifolds. The flat spacetime tetrad%
\begin{equation}
\begin{array}{l}
L^{\mu }\partial _{\mu }=\partial _{t}+\partial _{r} \\ 
\\ 
N^{\mu }\partial _{\mu }=\frac{r^{2}+a^{2}}{2(r^{2}+a^{2}\cos ^{2}\theta )}%
\left( \partial _{t}-\partial _{r}+\frac{2a}{r^{2}+a^{2}}\partial _{\varphi
}\right) \\ 
\\ 
M^{\mu }\partial _{\mu }=\frac{1}{\sqrt{2}(r+ia\cos \theta )}\left( ia\sin
\theta \partial _{t}+\partial _{\theta }+\frac{i}{\sin \theta }\partial
_{\varphi }\right)%
\end{array}
\label{i13}
\end{equation}%
and its corresponding covariant form%
\begin{equation}
\begin{array}{l}
L_{\mu }dx^{\mu }=dt-dr-a\sin ^{2}\theta \ d\varphi \\ 
\\ 
N_{\mu }dx^{\mu }=\frac{r^{2}+a^{2}}{2(r^{2}+a^{2}\cos ^{2}\theta )}[dt+%
\frac{r^{2}+2a^{2}\cos ^{2}\theta -a^{2}}{r^{2}+a^{2}}dr-a\sin ^{2}\theta \
d\varphi ] \\ 
\\ 
M_{\mu }dx^{\mu }=\frac{-1}{\sqrt{2}(r+ia\cos \theta )}[-ia\sin \theta \
(dt-dr)+(r^{2}+a^{2}\cos ^{2}\theta )d\theta + \\ 
\qquad \qquad +i\sin \theta (r^{2}+a^{2})d\varphi ]%
\end{array}
\label{i14}
\end{equation}%
determine a LCR-structure. This definition implies that the structure is
singular on the cylinder $r=0,\ \theta =\frac{\pi }{2}$. We will see below
that a more appropriate definition shows that this singularity is
artificial, depending on the structure coordinates of the ambient complex
manifold. Using the Kerr-Schild ansatz, we can find the following curved
spacetime LCR-structure%
\begin{equation}
\begin{array}{l}
\ell _{\mu }=L_{\mu }\quad ,\quad m_{\mu }=M_{\mu }\quad ,\quad n_{\mu
}=N_{\mu }+\frac{h(r)}{2(r^{2}+a^{2}\cos ^{2}\theta )}\ L_{\mu } \\ 
\\ 
\ell ^{\mu }=L^{\mu }\quad ,\quad m^{\mu }=M^{\mu }\quad ,\quad n^{\mu
}=N^{\mu }-\frac{h(r)}{2(r^{2}+a^{2}\cos ^{2}\theta )}\ L^{\mu } \\ 
\end{array}
\label{i15}
\end{equation}%
where $h(r)$ is an arbitrary function, which gives a regular spacetime. I
will call it Kerr-Newman LCR-structure because $h(r)=-2mr+e^{2}$ gives the
Kerr-Newman spacetime. The structure coordinates of the curved complex
structure are

\begin{equation}
\begin{array}{l}
z^{0}=t-r+ia\cos \theta -ia\quad ,\quad z^{1}=e^{i\varphi }\tan \frac{\theta 
}{2} \\ 
\\ 
z^{\widetilde{0}}=t+r-ia\cos \theta +ia-2f_{1}\quad ,\quad z^{\widetilde{1}%
}=-\frac{r+ia}{r-ia}\ e^{2iaf_{2}}\ e^{-i\varphi }\tan \frac{\theta }{2}%
\end{array}
\label{i16}
\end{equation}%
where the two new functions are

\begin{equation}
f_{1}(r)=\int \frac{h}{r^{2}+a^{2}+h}\ dr\quad ,\quad f_{2}(r)=\int \frac{h}{%
(r^{2}+a^{2}+h)(r^{2}+a^{2})}\ dr  \label{i17}
\end{equation}

The following relations give the Newman-Penrose spin coefficients%
\begin{equation}
\begin{tabular}{|l|}
\hline
$\alpha =\frac{1}{4}[(\ell n\partial \overline{m})+(\ell \overline{m}%
\partial n)-(n\overline{m}\partial \ell )-2(m\overline{m}\partial \overline{m%
})]$ \\ \hline
$\beta =\frac{1}{4}[(\ell n\partial m)+(\ell m\partial n)-(nm\partial \ell
)-2(m\overline{m}\partial m)]$ \\ \hline
$\gamma =\frac{1}{4}[(nm\partial \overline{m})-(n\overline{m}\partial m)-(m%
\overline{m}\partial n)+2(\ell n\partial n)]$ \\ \hline
$\varepsilon =\frac{1}{4}[(\ell m\partial \overline{m})-(\ell \overline{m}%
\partial m)-(m\overline{m}\partial \ell )+2(\ell n\partial \ell )]$ \\ \hline
$\mu =-\frac{1}{2}[(m\overline{m}\partial n)+(nm\partial \overline{m})+(n%
\overline{m}\partial m)]$ \\ \hline
$\pi =\frac{1}{2}[(\ell n\partial \overline{m})-(n\overline{m}\partial \ell
)-(\ell \overline{m}\partial n)]$ \\ \hline
$\rho =\frac{1}{2}[(\ell \overline{m}\partial m)+(\ell m\partial \overline{m}%
)-(m\overline{m}\partial \ell )]$ \\ \hline
$\tau =\frac{1}{2}[(nm\partial \ell )+(\ell m\partial n)+(\ell n\partial m)]$
\\ \hline
$\kappa =(\ell m\partial \ell )\quad ,\quad \sigma =(\ell m\partial m)$ \\ 
\hline
$\nu =-(n\overline{m}\partial n)\quad ,\quad \lambda =-(n\overline{m}%
\partial \overline{m})$ \\ \hline
\end{tabular}
\label{i18}
\end{equation}%
where the symbols $(...)$ are constructed according to the rule of the
following example $(\ell m\partial n)=(\ell ^{\mu }m^{\nu }-\ell ^{\nu
}m^{\mu })(\partial _{\mu }n_{\nu })$. These symbols can be directly
computed 
\begin{equation}
\begin{array}{l}
(\ell n\partial \ell )=0\quad ,\quad (\ell m\partial \ell )=0\quad ,\quad
(nm\partial \ell )=\frac{\sqrt{2}a^{2}\cos \theta \ \sin \theta }{(r+ia\cos
\theta )(r^{2}+a^{2}\cos ^{2}\theta )} \\ 
(m\overline{m}\partial \ell )=\frac{2ia\cos \theta }{(r^{2}+a^{2}\cos
^{2}\theta )}\quad ,\quad (\ell n\partial n)=\frac{-2ra^{2}\sin ^{2}\theta
^{\prime }-2rh+(r^{2}+a^{2}\cos ^{2}\theta )h}{2(r^{2}+a^{2}\cos ^{2}\theta
)^{2}} \\ 
(\ell m\partial n)=0\quad ,\quad (nm\partial n)=0\quad ,\quad (m\overline{m}%
\partial n)=\frac{i(r^{2}+a^{2}+h)a\cos \theta }{(r^{2}+a^{2}\cos ^{2}\theta
)^{2}} \\ 
(\ell n\partial m)=\frac{-i\sqrt{2}ar\sin \theta }{(r+ia\cos \theta
)(r^{2}+a^{2}\cos ^{2}\theta )}\quad ,\quad (\ell m\partial m)=0\quad ,\quad
(\ell \overline{m}\partial m)=\frac{-1}{r-ia\cos \theta } \\ 
(nm\partial m)=0\quad ,\quad (n\overline{m}\partial m)=\frac{r^{2}+a^{2}+h}{%
2(r-ia\cos \theta )(r^{2}+a^{2}\cos ^{2}\theta )} \\ 
(m\overline{m}\partial m)=-\frac{r\cos \theta +ia}{\sqrt{2}\sin \theta \
(r+ia\cos \theta )^{2}} \\ 
\end{array}
\label{i19}
\end{equation}

as the first step to the computation of the NP spin coefficients of this
null tetrad in the Lindquist coordinates%
\begin{equation}
\begin{tabular}{|l|}
\hline
$\alpha =\frac{ia(1+\sin ^{2}\theta )-r\cos \theta }{2\sqrt{2}\sin \theta \
(r-ia\cos \theta )^{2}}\quad ,\quad \beta =\frac{\cos \theta }{2\sqrt{2}\sin
\theta \ (r+ia\cos \theta )}$ \\ \hline
$\gamma =-\frac{a^{2}+iar\cos \theta +h}{2(r+ia\cos \theta )(r-ia\cos \theta
)^{2}}+\frac{h^{\prime }}{4(r+ia\cos \theta )(r-ia\cos \theta )}\quad ,\quad
\varepsilon =0$ \\ \hline
$\mu =-\frac{r^{2}+a^{2}+h}{2(r+ia\cos \theta )(r-ia\cos \theta )^{2}}\quad
,\quad \pi =\frac{ia\sin \theta }{\sqrt{2}(r-ia\cos \theta )^{2}}$ \\ \hline
$\rho =-\frac{1}{r-ia\cos \theta }\quad ,\quad \tau =-\frac{ia\sin \theta }{%
\sqrt{2}(r+ia\cos \theta )(r-ia\cos \theta )}$ \\ \hline
$\kappa =0\quad ,\quad \sigma =0\quad ,\quad \nu =0\quad ,\quad \lambda =0$
\\ \hline
\end{tabular}
\label{i20}
\end{equation}%
We will use the Kerr-Newman LCR manifold as an example for a better
understanding of the properties of LCR-manifolds.

\section{RELATIVE\ INVARIANTS OF LCR-STRUCTURES}

\setcounter{equation}{0}

Applying the conditions $\kappa =\sigma =\lambda =\nu =0$ to the relations (%
\ref{i5}) we find that the LCR-structure is determined by 
\begin{equation}
\begin{array}{l}
d\ell =Z_{1}\wedge \ell +(\rho -\overline{\rho })m\wedge \overline{m} \\ 
\\ 
dn=Z_{2}\wedge n+(\mu -\overline{\mu })m\wedge \overline{m} \\ 
\\ 
dm=Z\wedge m-(\tau +\overline{\pi })\ell \wedge n \\ 
\end{array}
\label{r1}
\end{equation}%
where \ 
\begin{equation}
\begin{array}{l}
Z_{1\mu }=(\theta _{1}+\mu +\overline{\mu })\ell _{\mu }+(\varepsilon +%
\overline{\varepsilon })n_{\mu }-(\alpha +\overline{\beta }-\overline{\tau }%
)m_{\mu }- \\ 
\qquad \quad -(\overline{\alpha }+\beta -\tau )\overline{m}_{\mu } \\ 
\\ 
Z_{2\mu }=-(\gamma +\overline{\gamma })\ell _{\mu }+(\theta _{2}-\rho -%
\overline{\rho })n_{\mu }-(\pi -\alpha -\overline{\beta })m_{\mu }- \\ 
\qquad \quad -(\overline{\pi }-\overline{\alpha }-\beta )\overline{m}_{\mu }
\\ 
\\ 
Z_{\mu }=(\gamma -\overline{\gamma }+\overline{\mu })\ell _{\mu
}+(\varepsilon -\overline{\varepsilon }-\rho )n_{\mu }-(\theta _{3}+\pi -%
\overline{\tau })m_{\mu }- \\ 
\qquad \quad -(\beta -\overline{\alpha })\overline{m}_{\mu } \\ 
\end{array}
\label{r2}
\end{equation}%
with the functions $\theta _{1}\ ,\ \theta _{2}\ ,\ \theta _{3}$\ \'{a}
priori arbitrary. Because of this particular form (\ref{r1}) we can
complexify the coordinates, apply (holomorphic) Frobenius theorem and define
the structure coordinates in the case of real analytic functions.

The integrability conditions (\ref{i6}) of the LCR-structure are invariant
under the transformations%
\begin{equation}
\begin{tabular}{l}
$\ell _{\mu }^{\prime }=\Lambda \ell _{\mu }\quad ,\quad \ell ^{\prime \mu }=%
\frac{1}{N}\ell ^{\mu }$ \\ 
\\ 
$n_{\mu }^{\prime }=Nn_{\mu }\quad ,\quad n^{\prime \mu }=\frac{1}{\Lambda }%
n^{\mu }$ \\ 
\\ 
$m_{\mu }^{\prime }=Mm_{\mu }\quad ,\quad m^{\prime \mu }=\frac{1}{\overline{%
M}}m^{\mu }$ \\ 
\end{tabular}
\label{r3}
\end{equation}%
which we will call tetrad-Weyl transformations. Under these transformations
the NP spin coefficients transform as follows%
\begin{equation}
\begin{tabular}{l}
$\alpha ^{\prime }=\frac{1}{M}\alpha +\frac{M\ \overline{M}-\Lambda N}{%
4M\Lambda N}(\overline{\tau }+\pi )+\frac{1}{4M}\overline{\delta }\ln \frac{%
\Lambda }{N\overline{M}^{2}}$ \\ 
$\beta ^{\prime }=\frac{1}{\overline{M}}\beta +\frac{M\ \overline{M}-\Lambda
N}{4\overline{M}\Lambda N}(\tau +\overline{\pi })+\frac{1}{4\overline{M}}%
\delta \ln \frac{\Lambda M^{2}}{N}$ \\ 
$\gamma ^{\prime }=\frac{1}{\Lambda }\gamma +\frac{M\ \overline{M}-\Lambda N%
}{4M\ \overline{M}\Lambda }(\overline{\mu }-\mu )+\frac{1}{4\Lambda }\Delta
\ln \frac{M}{N^{2}\overline{M}}$ \\ 
$\varepsilon ^{\prime }=\frac{1}{N}\varepsilon +\frac{M\ \overline{M}%
-\Lambda N}{4M\ \overline{M}N}(\overline{\rho }-\rho )+\frac{1}{4N}D\ln 
\frac{M\Lambda ^{2}}{\overline{M}}$ \\ 
$\mu ^{\prime }=\frac{1}{2\Lambda }(\mu +\overline{\mu })+\frac{N}{2M\ 
\overline{M}}(\mu -\overline{\mu })+\frac{1}{2\Lambda }\Delta \ln (M\ 
\overline{M})$ \\ 
$\rho ^{\prime }=\frac{1}{2N}(\rho +\overline{\rho })+\frac{\Lambda }{2M\ 
\overline{M}}(\rho -\overline{\rho })-\frac{1}{2N}D\ln (M\ \overline{M})$ \\ 
$\pi ^{\prime }=\frac{\overline{M}}{2\Lambda N}(\pi +\overline{\tau })+\frac{%
1}{2M}(\pi -\overline{\tau })+\frac{1}{2M}\overline{\delta }\ln (\Lambda N)$
\\ 
$\tau ^{\prime }=\frac{M}{2\Lambda N}(\tau +\overline{\pi })+\frac{1}{2%
\overline{M}}(\tau -\overline{\pi })-\frac{1}{2\overline{M}}\delta \ln
(\Lambda N)$ \\ 
$\kappa ^{\prime }=\frac{\Lambda }{N\overline{M}}\kappa \quad ,\quad \sigma
^{\prime }=\frac{M}{N\overline{M}}\sigma $ \\ 
$\nu ^{\prime }=\frac{N}{\Lambda M}\nu \quad ,\quad \lambda ^{\prime }=\frac{%
\overline{M}}{\Lambda M}\lambda $%
\end{tabular}
\label{r4}
\end{equation}%
We see that the following relations 
\begin{equation}
\begin{tabular}{l}
$\rho ^{\prime }-\overline{\rho ^{\prime }}=\frac{\Lambda }{M\overline{M}}%
(\rho -\overline{\rho })$ \\ 
$\mu ^{\prime }-\overline{\mu ^{\prime }}=\frac{N}{M\overline{M}}(\mu -%
\overline{\mu })$ \\ 
$\tau ^{\prime }+\overline{\pi ^{\prime }}=\frac{M}{\Lambda N}(\tau +%
\overline{\pi })$%
\end{tabular}
\label{r5}
\end{equation}%
establish the corresponding quantities as relative invariants of the
LCR-structure. That is, the LCR-structures are characterized by the
annihilation or not of these three quantities. A LCR-structure with
vanishing one of these three quantities is not equivalent with a
LCR-structure with non-vanishing the same quantity. In the generic case of
LCR-structures with non-vanishing all these three quantities, we can always
define the unique normalized tetrad ($\underline{\ell }\ ,\ \underline{n}\
,\ \underline{m}\ ,\ \overline{\underline{m}}$) which has 
\begin{equation}
\begin{tabular}{l}
$\rho -\overline{\rho }=i\quad ,\quad \mu -\overline{\mu }=-i\quad ,\quad
\tau +\overline{\pi }=-1$ \\ 
\end{tabular}
\label{r6}
\end{equation}%
These LCR-structures will be called generic and the corresponding unique
tetrad normalized.

The Kerr-Newman LCR-structure has 
\begin{equation}
\begin{tabular}{l}
$\rho -\overline{\rho }=\frac{-2ia\cos \theta }{(r+ia\cos \theta )(r-ia\cos
\theta )}$ \\ 
$\mu -\overline{\mu }=\frac{ia(r^{2}+a^{2}+h)\cos \theta }{(r+ia\cos \theta
)^{2}(r-ia\cos \theta )^{2}}$ \\ 
$\tau +\overline{\pi }=-\frac{i\sqrt{2}ar\sin \theta }{(r+ia\cos \theta
)^{2}(r-ia\cos \theta )}$%
\end{tabular}
\label{r7}
\end{equation}%
If $a\neq 0$ it has non-vanishing all the three relative invariants.

Notice that the vector fields $Z_{1\mu }\ ,\ Z_{2\mu }\ ,\ Z_{\mu }$
transform as gauge fields under the structure preserving transformations (%
\ref{r3}) \ 
\begin{equation}
\begin{array}{l}
Z_{1\mu }^{\prime }=Z_{1\mu }+\partial _{\mu }\ln \Lambda \quad ,\quad
Z_{2\mu }^{\prime }=Z_{2\mu }+\partial _{\mu }\ln N \\ 
\\ 
Z_{\mu }^{\prime }=Z_{\mu }+\partial _{\mu }\ln M \\ 
\end{array}
\label{r8}
\end{equation}%
Hence their differential forms \ 
\begin{equation}
\begin{array}{l}
F_{1}=dZ_{1}\quad ,\quad F_{2}=dZ_{2}\quad ,\quad F=dZ \\ 
\end{array}
\label{r9}
\end{equation}%
are LCR invariant. In the generic case of non vanishing relative invariants,
the gauge transformations are satisfied if \ 
\begin{equation}
\begin{array}{l}
\theta _{1}=n^{\mu }\partial _{\mu }\ln \frac{\rho -\overline{\rho }}{i}%
\quad ,\quad \theta _{2}=\ell ^{\mu }\partial _{\mu }\ln \frac{\mu -%
\overline{\mu }}{-i}\quad ,\quad \theta _{3}=\overline{m}^{\mu }\partial
_{\mu }\ln (-\tau -\overline{\pi }) \\ 
\end{array}
\label{r10}
\end{equation}

In the case of the Kerr-Newman LCR-manifold I find \ 
\begin{equation}
\begin{array}{l}
Z_{1\mu }=\frac{\sqrt{2}a^{2}\sin \theta \cos \theta }{(r+ia\cos \theta
)(r-ia\cos \theta )^{2}}m_{\mu }+\frac{\sqrt{2}a^{2}\sin \theta \cos \theta 
}{(r+ia\cos \theta )^{2}(r-ia\cos \theta )}\overline{m}_{\mu } \\ 
\\ 
Z_{2\mu }=(\frac{a^{2}r\sin ^{2}\theta +rh}{(r+ia\cos \theta )^{2}(r-ia\cos
\theta )^{2}}-\frac{h^{\prime }}{2(r+ia\cos \theta )(r-ia\cos \theta )})\ell
_{\mu }+ \\ 
\qquad \quad +(\frac{2r+h^{\prime }}{r^{2}+a^{2}+h}-\frac{2r}{(r+ia\cos
\theta )(r-ia\cos \theta )})n_{\mu } \\ 
\\ 
Z_{\mu }=-\frac{a^{2}+r^{2}+h}{2(r+ia\cos \theta )(r-ia\cos \theta )^{2}}%
\ell _{\mu }+\frac{1}{r-ia\cos \theta }n_{\mu }-\frac{r\cos \theta +ia}{%
\sqrt{2}\sin \theta (r+ia\cos \theta )(r-ia\cos \theta )}m_{\mu }- \\ 
\qquad \quad -\frac{r\cos \theta +ia}{\sqrt{2}\sin \theta (r+ia\cos \theta
)^{2}}\overline{m}_{\mu } \\ 
\end{array}
\label{r11}
\end{equation}%
and the corresponding tetrad-Weyl invariants are the differential forms \ 
\begin{equation}
\begin{array}{l}
F_{1}=dZ_{1}=\frac{4ra^{2}\sin \theta \cos \theta }{(r^{2}+a^{2}\cos \theta
)^{2}}dr\wedge d\theta \\ 
\\ 
F_{2}=dZ_{2}=\frac{4ra^{2}\sin \theta \cos \theta }{(r^{2}+a^{2}\cos \theta
)^{2}}dr\wedge d\theta \\ 
\\ 
F=dZ=-\frac{4ra^{2}\sin \theta \cos \theta }{(r^{2}+a^{2}\cos \theta )^{2}}%
dr\wedge d\theta \\ 
\end{array}%
\end{equation}%
which do not depend on $h(r)$. This may be an indication that the "mass" and
the "charge" of the Kerr-Newman LCR-manifold may not be of local origin.

Notice that the static spherically symmetric spacetimes have vanishing
relative invariants. Hence their LCR-structure is trivial.

I will now show that the LCR-structures with $\rho -\overline{\rho }\neq
0\neq \mu -\overline{\mu }$\ can be Cartan osculated with the $U(2)$\ group.

The structure equations of the $U(2)$\ group take the following appropriate
form \ 
\begin{equation}
\begin{array}{l}
\omega =U^{-1}dU=i%
\begin{pmatrix}
\ell & \overline{m} \\ 
m & n%
\end{pmatrix}%
\quad ,\quad d\omega +\omega \wedge \omega =0 \\ 
\\ 
d\ell =im\wedge \overline{m}\quad ,\quad dn=-im\wedge \overline{m}\quad
,\quad dm=i(\ell -n)\wedge m \\ 
\end{array}
\label{r12}
\end{equation}%
\bigskip

Notice that it is a LCR-manifold with vanishing third relative invariant $%
\tau +\overline{\pi }=0$. Hence the non-degenerate expression (\ref{r1}) may
be viewed as a Cartan $U(2)$\ osculation of the LCR-structure. Then the $%
U(2) $-Cartan curvature\ of a general LCR-structure in the (\ref{r6})
tetrad-Weyl condition is \ 
\begin{equation}
\begin{array}{l}
\Omega =i%
\begin{pmatrix}
Z_{1}\wedge \underline{\ell } & (\overline{Z}+i\underline{\ell }-i\underline{%
n})\wedge \overline{\underline{m}}+\underline{\ell }\wedge \underline{n} \\ 
(Z-i\underline{\ell }+i\underline{n})\wedge \underline{m}+\underline{\ell }%
\wedge \underline{n} & Z_{2}\wedge \underline{n}%
\end{pmatrix}
\\ 
\end{array}
\label{r13}
\end{equation}

\section{WEYL CURVATURE\ CLASSIFICATION OF LCR-STRUCTURES}

\setcounter{equation}{0}

We already know that the LCR-structure does not uniquely determine the
tetrad ($\ell \ ,\ n\ ,\ m\ ,\ \overline{m}$). It is defined up to a
tetrad-Weyl transformation (\ref{r3}). Hence the following real lorentzian
metric

\begin{equation}
\begin{array}{l}
g_{\mu \nu }=\ell _{\mu }n_{\nu }+n_{\mu }\ell _{\nu }-m_{{}\mu }\overline{m}%
_{\nu }-\overline{m}_{\mu }m_{\nu } \\ 
\end{array}
\label{w1}
\end{equation}%
is not uniquely determined. But the inverse problem of "how many
LCR-structures are compatible with a regular lorentzian riemannian
manifold?", finds a very interesting solution.

Taking into account that the integrability conditions of a LCR-structure
coincide with the existence of two geodetic and shear free null congruences $%
\ell ^{\mu }$ and $n^{\mu }$, the problem turns to that of finding the
geodetic and shear free null congruences of a metric $g_{\mu \nu }$. But not
all the metrics admit geodetic and shear free null congruences. Hence not
all the metrics are compatible with a LCR-structure.

\textbf{Definition: }If a null tetrad of a metric $g_{\mu \nu }$ satisfies
the LCR-structure conditions $\kappa =\sigma =\lambda =\nu =0$\ , we will
say that the corresponding lorentzian riemannian manifold admits a
LCR-structure.

It is well known that the null tetrads of a metric $g_{\mu \nu }$ transform
between each other with a Lorentz transformation\cite{CHAND}. We will always
consider regular metrics up to an appropriate coordinate transformation.
These metrics always admit at least one regular null tetrad up to an
appropriate coordinate transformation. This tetrad does not necessarily
determine geodetic and shear free congruences. But the existence of such a
regular metric and a regular null tetrad implies that all the scalar
quantities of the riemannian geometry are regular functions. The vector $%
\ell ^{\prime \mu }$ of a new null tetrad is related to ($\ell \ ,\ n\ ,\ m\
,\ \overline{m}$), with the relation

\begin{equation}
\begin{array}{l}
\ell ^{\prime \mu }=\ell ^{\mu }+\overline{b}m^{\mu }+b\overline{m}^{\mu }+b%
\overline{b}n^{\mu } \\ 
\end{array}
\label{w2}
\end{equation}%
where $b(x)$\ is a complex function. It is geodetic and shear free if $%
\kappa ^{\prime }=\sigma ^{\prime }=0$, which implies the algebraic relation%
\cite{CHAND}

\begin{equation}
\begin{array}{l}
\Psi _{0}^{\prime }=\Psi _{0}+4b\Psi _{1}+6b^{2}\Psi _{2}+4b^{3}\Psi
_{3}+b^{4}\Psi _{4}=0 \\ 
\end{array}
\label{w3}
\end{equation}%
The quantities $\Psi _{i}$\ are the Weyl scalars in the NP formalism
determined with the regular null tetrad. Therefore they must be regular
functions on the manifold. Hence we conclude that a geodetic and shear free
null congruence is determined by a four degree polynomial with regular
coefficients. Taking into account that a LCR-structure is determined by two
geodetic and shear free congruences, we conclude to the following four
cases\ \ 
\begin{equation}
\begin{array}{l}
Case\ I:\Psi _{1}\neq 0\ ,\ \Psi _{2}\neq 0\ ,\ \Psi _{3}\neq 0 \\ 
\\ 
Case\ II:\Psi _{1}\neq 0\ ,\ \Psi _{2}\neq 0\ ,\ \Psi _{3}=0 \\ 
\\ 
Case\ III:\Psi _{1}\neq 0\ ,\ \Psi _{2}=0\ ,\ \Psi _{3}=0 \\ 
\\ 
Case\ D:\Psi _{1}=0\ ,\ \Psi _{2}\neq 0\ ,\ \Psi _{3}=0 \\ 
\end{array}
\label{w4}
\end{equation}%
where $\Psi _{i}\ ,\ i=1,2,3$\ are now the non-vanishing Weyl scalars
relative to the tetrad which defines the LCR-structure. If the LCR-structure
is compatible with a conformally flat metric no restriction/classification
is imposed. I want to point out that "conformal flatness" is not invariant
under the tetrad-Weyl transformation (\ref{r3}). Therefore this property
applies to the larger set of spacetimes which become flat after a
tetrad-Weyl transformation.

\section{FLAHERTY'S PSEUDO-COMPLEX STRUCTURE}

\setcounter{equation}{0}

Any real analytic function on a totally real submanifold can be extended to
a holomorphic function on the ambient complex manifold. Applying this
property, the metric (\ref{w1}) takes a special form. Using the structure
coordinates $(z^{\alpha },\;z^{\widetilde{\alpha }})$,\ $\alpha =0,\ 1$ ,
the compatible metric $g_{\mu \nu }$ takes the form 
\begin{equation}
\begin{tabular}{l}
$ds^{2}=g_{\mu \nu }dx^{\mu }dx^{\nu }=2g_{\alpha \widetilde{\beta }}\
dz^{\alpha }dz^{\widetilde{\beta }}|_{M}$ \\ 
\end{tabular}
\label{f1}
\end{equation}%
where $g_{\alpha \widetilde{\beta }}$ are holomorphic functions of the
structure coordinates and the last term is taken on the LCR-manifold. Its
extension in the ambient complex manifold reveals the invariance of the
metric under the transformation $z^{\prime \alpha }=f^{\alpha }(z^{\beta
}),\ z^{\prime \widetilde{\alpha }}=f^{\widetilde{\alpha }}(z^{\widetilde{%
\beta }})$.

In order to describe these covariant transformations, Flaherty introduced%
\cite{FLAHE} the integrable pseudo-complex (hermitian) structure

\begin{equation}
\begin{array}{l}
J_{\mu }^{\;\nu }=i(\ell _{\mu }n^{\nu }-n_{\mu }\ell ^{\nu }-m_{\mu }%
\overline{m}^{\nu }+\overline{m}_{\mu }m^{\nu }) \\ 
\\ 
J_{\mu }^{\;\rho }J_{\rho }^{\;\nu }=-\delta _{\mu }^{\nu }\quad ,\quad
J_{\mu }^{\;\rho }J_{\nu }^{\;\sigma }g_{\rho \sigma }=g_{\mu \nu }%
\end{array}
\label{f2}
\end{equation}%
Notice that this complex structure on $M$, is not a real tensor like the
ordinary complex structure. In this presentation we will not use this
notion. We will stick on the more conventional notion of the CR structure
and I will consider the complex extensions of the metric to the holomorphic
metric $g_{ab}$ and of the pseudo-complex structure to the holomorphic
tensor $J_{a}^{\;b}$ , defined in the ambient complex manifold. In fact, $%
J_{a}^{\;b}$ is a second integrable complex structure, which commutes with
the original complex structure of the ambient complex manifold. The
LCR-structure preserving transformations respect both complex structures of
the ambient complex manifold. I will not continue into these formalities. I
simply want to point out that if we find a way to fix the metric, then the
connection $\gamma _{bc}^{a}$ of this pseudo-complex structure is an
adequate process to provide a simple classification of the LCR-structures.
The non vanishing components of this connection are 
\begin{equation}
\begin{tabular}{l}
$\gamma _{\beta \gamma }^{\alpha }=g^{\alpha \widetilde{\alpha }}\ \partial
_{\beta }g_{\gamma \widetilde{\alpha }}\quad ,\quad \gamma _{\widetilde{%
\beta }\widetilde{\gamma }}^{\widetilde{\alpha }}=g^{\alpha \widetilde{%
\alpha }}\ \partial _{\widetilde{\beta }}g_{\alpha \widetilde{\gamma }}$ \\ 
\end{tabular}
\label{f3}
\end{equation}%
The covariant derivative is defined as usual 
\begin{equation}
\begin{tabular}{l}
$\mathcal{D}_{a}T^{b}=\partial _{a}T^{b}+\gamma _{ac}^{b}T^{c}\quad ,\quad 
\mathcal{D}_{a}T_{b}=\partial _{a}T_{b}-\gamma _{ab}^{c}T_{c}$ \\ 
\end{tabular}
\label{f4}
\end{equation}%
and one can easily show that this covariant derivative annihilates the
metric, $\mathcal{D}_{a}g_{bc}=0$.

Flaherty has also showed that if the torsion of this connection $%
T_{ab}^{c}=\gamma _{ba}^{c}-\gamma _{ab}^{c}$ vanishes, the complex
structure is kaehlerian, $d(J_{\mu \nu }\ dx^{\mu }\wedge dx^{\nu })=0$, and
the vectors of the null tetrad are hypersurface orthogonal. This means that
the LCR-structure is trivial.

In the case of generic LCR-structures, we may use these covariant
derivatives to formulate the conventional classification scheme. We first
write in structure coordinates of both complex structures the unique generic
extended metric $\underline{g}_{ab}$ created from the unique generic tetrad
implied by fixing the relative invariants as in (\ref{r6}). We compute the
Flaherty connection, the corresponding torsion and curvature. Then we create
all their covariant derivatives. All the scalars one may construct with the
multiplication of these covariant tensors, properly contracted using the
generic metric, are invariant relative to the LCR-structure preserving
transformations. These scalars permit us to distinguish non-equivalent
LCR-structures.

\section{A\ PAIR\ OF\ 3-DIMENSIONAL\ CR\ STRUCTURES}

\setcounter{equation}{0}

From the forms of the four real relations (\ref{i9}), which define the
totally real submanifold of an 8 (real) dimensional complex manifold, we see
that the LCR-structure contains two 3-dimensional CR manifolds of the
hypersurface type, which have been extensively studied\cite{JACO1990}. We
will denote $\mathcal{I}^{+}$ the 3-dimensional submanifold determined by
the condition $\rho _{11}(\overline{z^{\alpha }},z^{\alpha })=0$ and $%
\mathcal{I}^{-}$ the 3-dimensional submanifold determined by the condition$\
\rho _{22}\left( \overline{z^{\widetilde{\alpha }}},z^{\widetilde{\alpha }%
}\right) =0$. We see that the complex tangent spaces of these two CR
manifolds coincide and they are spanned by $m^{\mu }\partial _{\mu }$ and $%
\overline{m}^{\mu }\partial _{\mu }$ . This coincidence relation is implied
by the complex condition $\rho _{12}\left( \overline{z^{\alpha }},z^{%
\widetilde{\alpha }}\right) =0$.

The classification of the 3-dimensional CR manifolds of the hypersurface
type have been extensively studied\cite{JACO1990}, using either the Moser or
the Cartan approach. In the present section I will study the importance of
these two 3-dimensional CR manifolds in the context of the LCR-structure,
considering known these two fundamental classification schemes.

We start from a 3-dimensional CR structure $\rho _{11}(\overline{z^{\alpha }}%
,z^{\alpha })=0$ and we build the family of the CR structures implied by a
parameter ($w$) dependent antiholomorphic transformation $z^{\prime 
\widetilde{\alpha }}=f^{\widetilde{\alpha }}(\overline{z^{\beta }};w)$. Then
removing the generic parameter $w$, we find a complex relation $\rho
_{12}\left( \overline{z^{\alpha }},z^{\widetilde{\alpha }}\right) =0$
between the structure coordinates. Hence the final output is the following
LCR-structure \ 
\begin{equation}
\begin{array}{l}
z^{\widetilde{\alpha }}=f^{\widetilde{\alpha }}\left( \overline{z^{\beta }}%
;w\right) \\ 
\\ 
\rho _{22}\left( \overline{z^{\widetilde{\alpha }}},z^{\widetilde{\alpha }%
}\right) =\rho _{22}\left( \overline{f^{\widetilde{\alpha }}}(z^{\beta }),f^{%
\widetilde{\alpha }}(\overline{z^{\beta }})\right) =\rho _{11}\left( 
\overline{z^{\alpha }},z^{\alpha }\right) =0 \\ 
\\ 
\rho _{12}\left( \overline{z^{\alpha }},z^{\widetilde{\alpha }}\right) =0 \\ 
\end{array}
\label{h1}
\end{equation}%
We see that this kind of LCR-structures have antiholomorphically equivalent $%
\mathcal{I}^{+}$ and $\mathcal{I}^{-}$\ 3-dimensional submanifolds.

It is well known that we can find complex coordinates $z^{0}=u+iU$\ and $z^{%
\widetilde{0}}=v+iV$ , such that the embedding real functions of the
3-dimensional CR submanifolds take the form\ \ 
\begin{equation}
\begin{array}{l}
i\left( \overline{z^{0}}-z^{0}\right) -2U\left( u,z^{1},\overline{z^{1}}%
\right) =0 \\ 
\\ 
i\left( \overline{z^{\widetilde{0}}}-z^{\widetilde{0}}\right) -2V\left( v,z^{%
\widetilde{1}},\overline{z^{\widetilde{1}}}\right) =0 \\ 
\end{array}
\label{h3}
\end{equation}%
Then we can define two characteristic coordinate systems of the embedable
LCR-manifold. The ($u,v,z^{1},\overline{z^{1}}$) implied by the (Frobenius)
submanifold $\mathcal{I}^{+}$ and ($u,v,z^{\widetilde{1}},\overline{z^{%
\widetilde{1}}}$) implied by $\mathcal{I}^{-}$.

The pseudoconvex 3-dimensional CR structures are usually osculated with the $%
SU(1,2)$ symmetric hyperquadric. But the LCR-structures with ($\rho -%
\overline{\rho }\neq 0\neq \mu -\overline{\mu }$) cannot be osculated by the 
$SU(1,2)$ symmetric surface\ \ 
\begin{equation}
\begin{array}{l}
\begin{pmatrix}
\overline{\zeta ^{01}} & \overline{\zeta ^{11}} & \overline{\zeta ^{21}} \\ 
\overline{\zeta ^{02}} & \overline{\zeta ^{12}} & \overline{\zeta ^{22}}%
\end{pmatrix}%
\begin{pmatrix}
0 & 0 & -2i \\ 
0 & -1 & 0 \\ 
2i & 0 & 0%
\end{pmatrix}%
\begin{pmatrix}
\zeta ^{01} & \zeta ^{02} \\ 
\zeta ^{11} & \zeta ^{12} \\ 
\zeta ^{21} & \zeta ^{22}%
\end{pmatrix}%
=0 \\ 
\\ 
z^{0}=\frac{\zeta ^{21}}{\zeta ^{01}}\quad ,\quad z^{1}=\frac{\zeta ^{11}}{%
\zeta ^{01}}\quad ,\quad z^{\widetilde{0}}=\frac{\zeta ^{22}}{\zeta ^{02}}%
\quad ,\quad z^{\widetilde{1}}=\frac{\zeta ^{12}}{\zeta ^{02}} \\ 
\end{array}
\label{h4}
\end{equation}%
which is not a LCR-manifold, because \ 
\begin{equation}
\begin{array}{l}
dz^{0}\wedge dz^{1}\wedge dz^{\widetilde{0}}\wedge dz^{\widetilde{1}}\mid
_{M}=0 \\ 
\end{array}
\label{h5}
\end{equation}%
We will see below that they are osculated by the characteristic boundary of
the $SU(2,2)$\ classical domain.

\section{SURFACES\ OF\ THE\ SU(2,2) CLASSICAL\ DOMAIN}

\setcounter{equation}{0}

In section II we found that a general spacetime admits at most four geodetic
and shear free null congruences. Therefore it can be compatible with a
limited number of LCR-structures. This limitation does not apply to
spacetimes which are flat up to a tetrad-Weyl transformation. From the
twistor formalism\cite{P-R1984} we know that the LCR-structures, which are
compatible with the Minkowski metric are given by the following embedding
functions%
\begin{equation}
\begin{array}{l}
\overline{X^{mi}}E_{mn}X^{nj}=0 \\ 
\\ 
K_{1}(X^{m1})=0=K_{2}(X^{m2})%
\end{array}
\label{d1}
\end{equation}%
where the $K_{i}(X^{mi})$ are homogeneous functions, which we will call Kerr
functions. Apparently, instead of considering two different Kerr functions,
one may consider two different points of the same algebraic surface of $%
CP^{3}$. The $4\times 4$ matrix $E_{mn}$ is the $SU(2,2)$\ preserving
matrix. These surfaces will be called flat LCR-structures. The $4\times 2$
matrix%
\begin{equation}
\begin{array}{l}
X=%
\begin{pmatrix}
X^{01} & X^{02} \\ 
X^{11} & X^{12} \\ 
X^{21} & X^{22} \\ 
X^{31} & X^{32}%
\end{pmatrix}
\\ 
\end{array}
\label{d2}
\end{equation}%
must have rank 2. Using the chiral representation of $E_{mn}$%
\begin{equation}
\begin{array}{l}
E=%
\begin{pmatrix}
0 & I \\ 
I & 0%
\end{pmatrix}
\\ 
\end{array}
\label{d3}
\end{equation}%
and the notation%
\begin{equation}
\begin{array}{l}
X=%
\begin{pmatrix}
\lambda \\ 
-ir\lambda%
\end{pmatrix}
\\ 
\end{array}
\label{d4}
\end{equation}%
for $\det \lambda \neq 0$, the first relation of (\ref{d1}) takes the form%
\begin{equation}
\begin{array}{l}
i(r^{\dagger }-r)=0 \\ 
\end{array}
\label{d5}
\end{equation}%
That is the LCR-manifold is fixed without using the Kerr functions. This
manifold is the characteristic (Shilov) boundary of the $SU(2,2)$ classical
domain or a part of it. The Kerr functions determine the LCR-structures as $%
CP^{3}$ sections on this precise manifold. They are in fact defined by two
different points of an algebraic surface of $CP^{3}$.

The following embedding functions

\begin{equation}
\begin{array}{l}
\overline{X^{mi}}E_{mn}X^{nj}=G_{ij}(\overline{X^{mi}},X^{mj}) \\ 
\\ 
K_{1}(X^{m1})=0=K_{2}(X^{m2})%
\end{array}
\label{d6}
\end{equation}%
provide a formal osculation of a general LCR-structure with the Shilov
boundary of the $SU(2,2)$ classical domain.

Using the following spinorial form of the rank-2 matrix $X^{mj}$ in its
unbounded realization 
\begin{equation}
\begin{array}{l}
X^{mj}=%
\begin{pmatrix}
\lambda ^{Aj} \\ 
-ir_{A^{\prime }B}\lambda ^{Bj}%
\end{pmatrix}
\\ 
\end{array}
\label{d7}
\end{equation}%
and the null tetrad 
\begin{equation}
\begin{array}{l}
L^{a}=\frac{1}{\sqrt{2}}\overline{\lambda }^{A^{\prime }1}\lambda
^{B1}\sigma _{A^{\prime }B}^{a}\quad ,\quad N^{a}=\frac{1}{\sqrt{2}}%
\overline{\lambda }^{A^{\prime }2}\lambda ^{B2}\sigma _{A^{\prime
}B}^{a}\quad ,\quad M^{a}=\frac{1}{\sqrt{2}}\overline{\lambda }^{A^{\prime
}2}\lambda ^{B1}\sigma _{A^{\prime }B}^{a} \\ 
\\ 
\epsilon _{AB}\lambda ^{A1}\lambda ^{B2}=1 \\ 
\end{array}
\label{d8}
\end{equation}%
the above relations (\ref{d6}) take the form 
\begin{equation}
\begin{array}{l}
\rho _{11}=2\sqrt{2}y^{a}L_{a}-G_{11}(\overline{Y^{m1}},Y^{n1})=0 \\ 
\\ 
\rho _{12}=2\sqrt{2}y^{a}\overline{M}_{a}-G_{12}(\overline{Y^{m1}},Y^{n2})=0
\\ 
\\ 
\rho _{22}=2\sqrt{2}y^{a}N_{a}-G_{22}(\overline{Y^{m2}},Y^{n2})=0%
\end{array}
\label{d9}
\end{equation}%
where $y^{a}$\ is the imaginary part of $r^{a}=x^{a}+iy^{a}$\ defined by the
relation $r_{A^{\prime }B}=r^{a}\sigma _{aA^{\prime }B}$\ and $\sigma
_{A^{\prime }B}^{a}$ being the identity and the three Pauli matrices. The
surface satisfies the relation \ 
\begin{equation}
\begin{array}{l}
y^{a}=\frac{1}{2\sqrt{2}}[G_{22}L^{a}+G_{11}N^{a}-G_{12}M^{a}-\overline{%
G_{12}}\overline{M}^{a}] \\ 
\end{array}
\label{d10}
\end{equation}%
which, combined with the computation of $\lambda ^{Ai}$\ as functions of $%
r^{a}$, using the Kerr conditions $K_{i}(X^{mi})$, permits us to compute $%
y^{a}=$ $h^{a}(x)$ as functions of the real part of $r^{a}$. In fact the
normal form\cite{BAOU} of any totally real submanifold of a complex manifold
is $y^{a}=y^{a}(x)$. The characteristic property of the LCR-manifolds is
that they can be considered as surfaces of the $SU(2,2)$ symmetric classical
domain with $y^{a}$ and $x^{a}$\ related to the projective coordinates of
the chiral representation. Besides, $y^{a}(x)$ must satisfy the relation \ 
\begin{equation}
\begin{array}{l}
\begin{pmatrix}
y^{a}L_{a} & y^{a}M_{a} \\ 
y^{a}\overline{M_{a}} & y^{a}N_{a}%
\end{pmatrix}%
=\frac{1}{2\sqrt{2}}%
\begin{pmatrix}
G_{11} & G_{12} \\ 
\overline{G_{12}} & G_{22}%
\end{pmatrix}
\\ 
\end{array}
\label{d11}
\end{equation}%
with the homogeneity factors of $G_{ij}$\ properly fixed.

Notice that this surface may always be assumed to belong into the "upper
half-plane", because the tetrad may be arranged such that $y^{0}>0$, but
this surface does not generally belong into the $SU(2,2)$ Siegel fundamental
domain, because \ 
\begin{equation}
\begin{array}{l}
y^{a}y^{b}\eta _{ab}=\frac{1}{4}[G_{11}G_{22}-G_{12}\overline{G_{12}}] \\ 
\end{array}
\label{d12}
\end{equation}%
is not always positive. The regular surfaces (with an upper bound) can
always be brought inside the Siegel domain (and its holomorphic bounded
classical domain) with an holomorphic complex time translation. 
\begin{equation}
\begin{array}{l}
\begin{pmatrix}
\lambda ^{\prime Aj} \\ 
-ir_{B^{\prime }B}^{\prime }\lambda ^{\prime Bj}%
\end{pmatrix}%
=%
\begin{pmatrix}
I & 0 \\ 
dI & I%
\end{pmatrix}%
\begin{pmatrix}
\lambda ^{Aj} \\ 
-ir_{B^{\prime }B}\lambda ^{Bj}%
\end{pmatrix}
\\ 
\end{array}
\label{d13}
\end{equation}%
which implies the transformation $y^{\prime a}=y^{a}+(d,0,0,0)$. An
appropriate constant $d$\ makes $y^{\prime 0}>0$\ and $y^{\prime a}y^{\prime
b}\eta _{ab}>0$. Apparently this constant $d$ is not uniquely determined.

The osculating form (\ref{d6}) preserves the $SU(2,2)$ group and vice-versa
the group preserves the LCR-structure. Hence the LCR-manifolds appear as
transitive sets of the $SU(2,2)$ group, which we will now denote as $\Sigma
\lbrack SU(2,2)]$. In a quantum field theoretic model, which is invariant
under the LCR-structure preserving transformations, the $\Sigma \lbrack
SU(2,2)]$ sets will appear as multiplets of the $SU(2,2)$ group, if the
group is not spontaneously broken. The exact Poincar\'{e} group is expected
to emerge after an appropriate spontaneous breaking of the $SU(2,2)$
symmetry. Therefore the sets $\Sigma \lbrack Poincar\acute{e}]$\ may present
special physical interest. Hence the LCR-manifolds, which are realizable as
surfaces in the $SU(2,2)$ classical domain are the "particles" of the model
and those which are not realizable (if they exist) are the "unparticles".

\subsection{Asymptotically flat LCR-manifolds}

The osculation of the LCR-manifold with the Shilov boundary of the $SU(2,2)$
classical domain, permits us to transfer the notion of "asymptotically flat
spacetimes at null infinities" into the terminology of LCR-structures. Using
the coordinates $t\ ,\ r$ , such that \ \ 
\begin{equation}
\begin{array}{l}
\frac{\overline{z^{0}}+z^{0}}{2}=u=t-r \\ 
\\ 
\frac{\overline{z^{\widetilde{0}}}-z^{\widetilde{0}}}{2}=v=t+r \\ 
\end{array}
\label{d14}
\end{equation}%
we may consider that the gravitational content vanishes in the two
limite-surfaces $\mathcal{J}^{+}=\{r\rightarrow \infty \ ,\ with\ z^{\alpha
}\ const\}$ and $\mathcal{J}^{-}=\{r\rightarrow \infty \ ,\ with\ z^{%
\widetilde{\alpha }}\ const\}$. It is achieved if $G_{11}=G_{22}=0$. Notice
that these surfaces are always outside the classical domain because (\ref%
{d12}) implies $y^{a}y^{b}\eta _{ab}<0$. The asymptotically flat CR
manifolds are invariant under the $SU(2,2)$ transformations, because they
preserve the conditions $G_{11}=G_{22}=0$.

The term "chiral" refers to the representation of $E$ as the $\gamma ^{0}$
Dirac matrix and should not be related (and it will not be related) to the
Dirac equation. The chiral representation of $E$ gives the unbounded
realization of the $SU(2,2)$ classical domain. Its bounded realization is
found using the following Dirac representation of the invariant matrix \ 
\begin{equation}
E=%
\begin{pmatrix}
I & 0 \\ 
0 & -I%
\end{pmatrix}
\label{d15}
\end{equation}%
where the projective coordinates $z$ are denoted by 
\begin{equation}
\begin{array}{l}
X^{mj}=%
\begin{pmatrix}
\lambda \\ 
z\lambda%
\end{pmatrix}
\\ 
\end{array}
\label{d16}
\end{equation}%
for $\det \lambda \neq 0$. The two projective coordinates are related with%
\begin{equation}
\begin{array}{l}
r=i(I+z)(I-z)^{-1}=i(I-z)^{-1}(I+z) \\ 
\\ 
z=(r-iI)(r+iI)^{-1}=(r+iI)^{-1}(r-iI)%
\end{array}
\label{d17}
\end{equation}

Notice that the "real axis" of the chiral representation does not cover all
the $U(2)$ Shilov boundary of the Dirac representation. The two boundaries $%
\mathfrak{J}^{\mathfrak{\pm }}$ are surfaces of $U(2)$. $\mathfrak{J}^{%
\mathfrak{+}}$ is\cite{P-R1984}\ $\tau +\rho =\pi \ ,\ (-\pi \leq \tau -\rho
\leq \pi )$, and $\mathfrak{J}^{\mathfrak{-}}$ is $\tau -\rho =\pi \ ,\
(-\pi \leq \tau +\rho \leq \pi )$. This fact permits us to distinguish the
LCR-manifolds into periodic and non-periodic. The periodic manifolds are
those permitting the identification $\mathfrak{J}^{\mathfrak{+}}=\mathfrak{J}%
^{\mathfrak{-}}$ and they will be physically interpreted as excitation
modes. The non-periodic manifolds are solitonic configurations according to
the general rule.

The LCR-structures of the Minkowski spacetime are periodic, because all the
null geodesics are periodic\cite{P-R1984}. The topology of the compactified
Minkowski spacetime turns out to be \textrm{M}$^{\mathrm{\#}}\sim
S^{3}\times S^{1}$.\ 

The Kerr-Newman LCR-manifold is a solitonic configuration because it is not
periodic. Around $\mathcal{J}^{\mathfrak{+}}$ the coordinates $(u,\ w=\frac{1%
}{r},\ \theta ,\ \varphi )$ are used, where the integrable tetrad takes the
asymptotic form

\begin{equation}
\begin{array}{l}
\ell \simeq \lbrack du-a\sin ^{2}\theta \ d\varphi ] \\ 
\\ 
n\simeq \lbrack w^{2}\ du-\ 2(1+2mw)dw-aw^{2}\sin ^{2}\theta \ d\varphi ] \\ 
\\ 
m\simeq \lbrack iaw^{2}\sin \theta \ du-(1+a^{2}w^{2}\cos ^{2}\theta )\
d\theta - \\ 
\qquad -i\sin \theta (1+a^{2}w^{2})\ d\varphi ]%
\end{array}
\label{d18}
\end{equation}%
The physical space is for $w>0$\ and the integrable tetrad is regular on $%
\mathcal{J}^{\mathfrak{+}}$ up to a factor, which does not affect the
congruences, and\ it can be regularly extended to $w<0$. Around $\mathcal{J}%
^{\mathfrak{-}}$ the coordinates $(v,\ w^{\prime },\ \theta ^{\prime },\
\varphi ^{\prime })$ are used with

\begin{equation}
\begin{array}{l}
dv\simeq du+2(1+\frac{2m}{r})\ dr \\ 
\\ 
dw^{\prime }\simeq -dw\quad ,\quad d\theta ^{\prime }\simeq d\theta \\ 
\\ 
d\varphi ^{\prime }\simeq d\varphi +\frac{2a}{r^{2}}(1+2mr)\ dr%
\end{array}
\label{d19}
\end{equation}%
and the integrable tetrad takes the form

\begin{equation}
\begin{array}{l}
\ell \simeq \lbrack w^{\prime 2}\ dv-2(1-2mw^{\prime })\ dw^{\prime
}-aw^{\prime 2}\sin ^{2}\theta ^{\prime }\ d\varphi ^{\prime }] \\ 
\\ 
n\simeq \lbrack dv-a\sin ^{2}\theta ^{\prime }\ d\varphi ^{\prime }] \\ 
\\ 
m\simeq \lbrack iaw^{\prime 2}\sin \theta \ dv-(1+a^{2}w^{\prime 2}\cos
^{2}\theta ^{\prime })\ d\theta ^{\prime }- \\ 
\qquad -i\sin \theta ^{\prime }(1+aw^{\prime 2})\ d\varphi ^{\prime }]%
\end{array}
\label{d20}
\end{equation}%
The physical space is for $w<0$\ and the integrable tetrad is regular on $%
\mathcal{J}^{\mathfrak{-}}$ up to a factor, which does not affect the
congruences, and\ it can be regularly extended to $w>0$. If the mass term
vanishes the two regions $\mathcal{J}^{\mathfrak{+}}$ and $\mathcal{JJ}^{%
\mathfrak{-}}$ can be identified\ and the $\ell ^{\mu }$ and $n^{\mu }$
congruences are interchanged, when $\mathcal{J}^{\mathfrak{+}}$ $(\equiv 
\mathcal{J}^{\mathfrak{-}})$ is crossed. When $m\neq 0$\ these two regions
cannot be identified\cite{P-R1984} and the LCR-structure cannot be extended
across $\mathcal{J}^{\mathfrak{+}}$ into $\mathcal{J}^{\mathfrak{-}}$,
permitting the identification these two boundaries.

Notice that if the mass of a solitonic LCR-manifold is defined with this
asymptotic procedure and $m>0$ , then it can be absorbed into the
dimensional coordinates and constants as follows

\begin{equation}
\begin{array}{l}
w=\frac{\widetilde{w}}{m}\quad ,\quad u=m\widetilde{u}\quad ,\quad a=m%
\widetilde{a} \\ 
\\ 
w^{\prime }=\frac{\widetilde{w^{\prime }}}{m}\quad ,\quad v=m\widetilde{v}
\\ 
\end{array}
\label{d21}
\end{equation}%
and it is factored out of the tetrad. This means that the mass parameter is
a relative invariant of the CR structure. That is only counts whether it
vanishes or not. Its precise value does not affect the structure.

\subsection{The Poincar\'{e} group}

The asymptotically flat LCR-submanifolds of the classical domain,
transformed to each other through an $SU(2,2)$ transformation, have the same
LCR-structure. Therefore we will say that they belong into a surface
representation of the $SU(2,2)$ group, which we already denoted $\Sigma
\lbrack SU(2,2)]$. On the other hand, the two boundaries $\mathcal{J}^{\pm }$
of the asymptotically flat LCR-manifolds are 3-dimensional surfaces of the
Shilov boundary, which meet at a point at spatial infinity. A general
theorem, valid for all classical domains, states that the automorphic
analytic transformations, which preserve a point of the characteristic
boundary in the bounded realization, become linear transformations in the
unbounded realization of the classical domain\cite{PIAT1966}. In the present
case of the $SU(2,2)$ classical domain these linear transformations form the 
$[Poincar\acute{e}]\times \lbrack dilation]$ subgroup. Therefore a vacuum
configuration with a singularity at a point in the Shilov boundary will
cause spontaneously breaking of the conformal group $SU(2,2)$ down to its $%
[Poincar\acute{e}]\times \lbrack dilation]$ subgroup.

The set of the transferred into the classical domain asymptotically flat
LCR-submanifolds, which transform between each other through a Poincar\'{e}
transformation have the normal form \ 
\begin{equation}
\begin{array}{l}
y^{a}=k^{a}+h^{a}(x) \\ 
\\ 
k^{a}k^{b}\eta _{ab}=m^{2} \\ 
\end{array}
\label{d22}
\end{equation}%
with the "momentum" $k^{a}$ characterizing each element of the set $\Sigma
\lbrack Poincar\acute{e}]$. The positive constant $m$ is fixed by the
condition \ 
\begin{equation}
\begin{array}{l}
m=\underset{\Sigma \lbrack P]}{\max }\left( \sqrt{(h^{i})^{2}}-h^{0}\right)
\\ 
\end{array}
\label{d23}
\end{equation}%
That is, the "mass" $m$ of an asymptotically flat set $\Sigma \lbrack Poincar%
\acute{e}]$ is defined to be the minimal value of $d$ of (\ref{d13}), that
transfers all the elements of $\Sigma \lbrack Poincar\acute{e}]$ inside the
classical domain.

The emergence of the Poincar\'{e} group is crucial for the effective
description of the dynamics of static solitons in the quantum cosmodynamic
model\cite{RAG2013}. The unitary representation of the set $\Sigma \lbrack
Poincar\acute{e}]$ is a free quantum field $\Psi (x)$ which satisfies the
scalar, spinorial or vector field equations depending on the spin of the
soliton. The quantum representation of the Kerr-Newman type soliton (\ref%
{i15}) has to satisfy the following Dirac equation \ 
\begin{equation}
\begin{array}{l}
\left( \gamma ^{a}(i\frac{\partial }{\partial x^{a}}-h_{a})-m\right) \Psi
(x)=0 \\ 
\end{array}
\label{d24}
\end{equation}%
because it is already known that its gyromagnetic ratio is fermionic. Notice
that the effective field representation $\Psi (x)$ incorporates the
solitonic form factor $h^{a}(x)$ through the above equation. In the
vanishing form factor approximation, the Kerr-Newman satisfies the ordinary
Dirac equation. Therefore the quantum cosmodynamic model may formally
generate current quantum field theories as effective theories. On the other
hand non-static LCR-manifolds with appropriate behavior at $t\rightarrow \pm
\infty $ describe the interaction of the asymptotic static LCR-manifolds.
Therefore it is very interesting to find all the regular static axially
symmetric LCR-manifolds.

\subsection{Complex trajectory emergence from LCR-manifolds}

Newman\cite{N2005} has shown that in each 3-dimensinal CR-manifold the Kerr
function may be replaced with a complex trajectory $\xi _{A^{\prime }B}(\tau
)=\xi _{b}(\tau )\sigma _{A^{\prime }B}^{b}$ , where $\tau $\ is a complex
parameter. In the present formalism, the Newman procedure is transcribed by
considering the following parametrization of $X^{mi}$\ 
\begin{equation}
X^{mi}=%
\begin{pmatrix}
\lambda ^{Ai} \\ 
-i\xi _{A^{\prime }B}^{i}(\tau _{i})\lambda ^{Bi}%
\end{pmatrix}
\label{d25}
\end{equation}%
Its combination with (\ref{d7}) implies that the complex parameters $\tau
_{i}$ are determined via the conditions $\det [r_{A^{\prime }B}-\xi
_{A^{\prime }B}^{i}(\tau _{i})]=0$ , which assure that the two linear
equations $[r_{A^{\prime }B}-\xi _{A^{\prime }B}^{i}(\tau _{i})]\lambda
^{Bi}=0$\ admit non-vanishing solutions. The variables $\tau _{1},\ \frac{%
\lambda ^{11}}{\lambda ^{01}},\ \tau _{2},\ \frac{\lambda ^{12}}{\lambda
^{02}}$ may be considered as structure coordinates of the embedable
LCR-structure. Hence the trajectories appear as moduli parameters of the
LCR-structures.

In the general case we need two complex trajectories for each LCR-structure.
But, notice that we may have one complex trajectory with two solutions. In
this possibility the LCR-surface of the Grassmannian manifold may be viewed
as a "moving particle" with an intrinsic complex component. One can easily
prove that the quadratic Kerr polynomial%
\begin{equation}
Z^{1}Z^{2}-Z^{0}Z^{3}+2aZ^{0}Z^{1}=0  \label{d26}
\end{equation}%
is implied by the trajectory $\xi ^{b}(\tau )=(\tau \ ,\ 0\ ,\ 0\ ,\ ia)$.

\section{EXAMPLES\ OF SYMMETRIC\ LCR-MANIFOLDS}

\setcounter{equation}{0}

In order to better understand the LCR-manifolds, we have to work with
special cases. We will work with the $SU(2,2)$\ osculation, which contains
the physically interesting Poincar\'{e} transformations, but analogous
approaches can be applied to the other subgroups of $SU(2,2)$.

In a quantum model, the LCR-structure solutions $X^{mi}(x)$ correspond to
quantum states. If the $SU(2,2)$\ group is not spontaneously broken, the
transformations of $X^{mi}(x)$ generate a group representation in the
corresponding Hilbert space of the states of LCR-manifolds. This
representation includes eigenstates of the three commuting generators of the 
$SU(2,2)$\ group. These eigenstates remain invariant under the
transformations along the corresponding generators. This implies that the
eigenstates "LCR-manifolds" will correspond to solutions of the symmetric
conditions (\ref{d6}). Therefore we will look for asymptotically flat
LCR-manifolds which satisfy conditions symmetric relative to
time-translation, rotation and scale transformations.

The time translation transformation is 
\begin{equation}
\delta X^{mi}=i\epsilon ^{0}[\mathrm{P}_{0}]_{n}^{m}X^{ni}  \label{e1}
\end{equation}%
where $\mathrm{P}_{\mu }=-\frac{1}{2}\gamma _{\mu }(1+\gamma _{5})$.\ In the
chiral representation, we have%
\begin{equation}
\begin{array}{l}
\delta X^{0i}=0\qquad ,\qquad \delta X^{1i}=0 \\ 
\\ 
\delta X^{2i}=-i\epsilon ^{0}X^{0i}\qquad ,\qquad \delta X^{3i}=-i\epsilon
^{0}X^{1i}%
\end{array}
\label{e2}
\end{equation}

The rotation transformations around the z-axis is

\begin{equation}
\delta X^{mi}=i\epsilon ^{12}[\mathrm{\Sigma }_{12}]_{n}^{m}X^{ni}
\label{e3}
\end{equation}%
where $\mathrm{\Sigma }_{\mu \nu }=\frac{1}{2}\sigma _{\mu \nu }=\frac{i}{4}%
(\gamma _{\mu }\gamma _{\nu }-\gamma _{\nu }\gamma _{\mu })$. Then we have%
\begin{equation}
\begin{array}{l}
\delta X^{0i}=-i\frac{\epsilon ^{12}}{2}X^{0i}\qquad ,\qquad \delta X^{1i}=i%
\frac{\epsilon ^{12}}{2}X^{1i} \\ 
\\ 
\delta X^{2i}=-i\frac{\epsilon ^{12}}{2}X^{2i}\qquad ,\qquad \delta X^{3i}=i%
\frac{\epsilon ^{12}}{2}X^{3i}%
\end{array}
\label{e4}
\end{equation}

Taking into account that the algebraic surfaces of $CP^{3}$ are locally
determined by polynomials (Chow's theorem), we will investigate quadratic
polynomial surfaces, which remain invariant under the above transformations.
The two solutions of the quadratic polynomial will determine the coordinates 
$z^{\alpha }$ and $z^{\widetilde{\alpha }}$\ respectively. I actually found
the following two invariant quadratic surfaces%
\begin{equation}
\begin{array}{l}
CASE\ I:\qquad Z^{1}Z^{2}-Z^{0}Z^{3}+2aZ^{0}Z^{1}=0 \\ 
\\ 
CASE\ II:\qquad Z^{0}Z^{1}=0%
\end{array}
\label{e5}
\end{equation}

I will first consider the surface of CASE I. In this case the convenient
structure coordinates are%
\begin{equation}
\begin{array}{l}
z^{0}=i\frac{X^{21}}{X^{01}}\quad ,\quad z^{1}=\frac{X^{11}}{X^{01}}\quad
,\quad z^{\widetilde{0}}=i\frac{X^{32}}{X^{12}}\quad ,\quad z^{\widetilde{1}%
}=-\frac{X^{02}}{X^{12}} \\ 
\end{array}
\label{e6}
\end{equation}%
which under time translation and z-rotation transform as follows%
\begin{equation}
\begin{array}{l}
\delta z^{0}=\epsilon ^{0}\qquad ,\qquad \delta z^{1}=0\qquad ,\qquad \delta
z^{\widetilde{0}}=\epsilon ^{0}\qquad ,\qquad \delta z^{\widetilde{1}}=0 \\ 
\\ 
\delta z^{0}=0\qquad ,\qquad \delta z^{1}=i\epsilon ^{12}z^{1}\qquad ,\qquad
\delta z^{\widetilde{0}}=0\qquad ,\qquad \delta z^{\widetilde{1}}=-i\epsilon
^{12}z^{\widetilde{1}}%
\end{array}
\label{e7}
\end{equation}%
I found the following invariant LCR-surface 
\begin{equation}
\begin{array}{l}
z^{0}=(t-r)+iU\qquad ,\qquad z^{\widetilde{0}}=(t+r)+iV \\ 
\\ 
U=-2a\frac{z^{1}\overline{z^{1}}}{1+z^{1}\overline{z^{1}}}\qquad ,\qquad V=2a%
\frac{z^{\widetilde{\widetilde{1}}}\overline{z^{\widetilde{\widetilde{1}}}}}{%
1+z^{\widetilde{\widetilde{1}}}\overline{z^{\widetilde{\widetilde{1}}}}} \\ 
\\ 
z^{\widetilde{1}}=\frac{r+ia}{r-ia}e^{2if(r)}\ \overline{z^{1}} \\ 
\end{array}
\label{e8}
\end{equation}%
where $f(r)$\ is an arbitrary real function, and $f(r)=0$ gives a flat
LCR-structure. These are not the most general invariant LCR-manifolds. The
additional condition $U+V=0$ has been assumed. A simple investigation shows
that this LCR-manifold is (in different coordinates) the static solution (%
\ref{i15}) found in section II using the Kerr-Schild ansatz. One may easily
compute the corresponding tetrad up to their arbitrary factors $N_{1}$, $%
N_{2}$ and $N_{3}$.%
\begin{equation}
\begin{array}{l}
\ell =N_{1}[dt-dr-a\sin ^{2}\theta \ d\varphi ] \\ 
\\ 
n=N_{2}[dt+(\frac{r^{2}+a^{2}\cos 2\theta }{r^{2}+a^{2}}-2a\sin ^{2}\theta \ 
\frac{df}{dr})dr-a\sin ^{2}\theta \ d\varphi ] \\ 
\\ 
m=N_{3}[-ia\sin \theta \ (dt-dr)+(r^{2}+a^{2}\cos ^{2}\theta )d\theta
+i(r^{2}+a^{2})\sin \theta d\varphi ] \\ 
\end{array}
\label{e9}
\end{equation}%
where $z^{1}=\tan \frac{\theta }{2}\ e^{i\varphi }$. The corresponding
projective coordinates $r^{a}$ are found using the relations%
\begin{equation}
\begin{array}{l}
r^{0}-r^{3}=r_{0^{\prime }0}=i\frac{X^{21}X^{12}-X^{11}X^{22}}{%
X^{01}X^{12}-X^{11}X^{02}}=\frac{z^{0}+(z^{\widetilde{0}}-2ia)z^{1}z^{%
\widetilde{1}}}{1+z^{1}z^{\widetilde{1}}} \\ 
\\ 
-r^{1}+ir^{3}=r_{0^{\prime }1}=i\frac{X^{01}X^{22}-X^{21}X^{02}}{%
X^{01}X^{12}-X^{11}X^{02}}=\frac{(z^{0}-z^{\widetilde{0}}+2ia)z^{\widetilde{1%
}}}{1+z^{1}z^{\widetilde{1}}} \\ 
\\ 
-r^{1}-ir^{3}=r_{1^{\prime }0}=i\frac{X^{31}X^{12}-X^{11}X^{32}}{%
X^{01}X^{12}-X^{11}X^{02}}=\frac{(z^{0}-z^{\widetilde{0}}+2ia)z^{1}}{%
1+z^{1}z^{\widetilde{1}}} \\ 
\\ 
r^{0}+r^{3}=r_{1^{\prime }1}=i\frac{X^{01}X^{32}-X^{31}X^{02}}{%
X^{01}X^{12}-X^{11}X^{02}}=\frac{z^{\widetilde{0}}+(z^{0}+2ia)z^{1}z^{%
\widetilde{1}}}{1+z^{1}z^{\widetilde{1}}} \\ 
\end{array}
\label{e10}
\end{equation}

The dilation transformation is

\begin{equation}
\delta X^{mi}=-i\epsilon \lbrack \mathrm{D}]_{n}^{m}X^{ni}  \label{e11}
\end{equation}%
with $\mathrm{D}=-\frac{i}{2}\gamma _{5}$. Then we have%
\begin{equation}
\begin{array}{l}
\delta X^{0i}=-\frac{\epsilon }{2}X^{0i}\qquad ,\qquad \delta X^{1i}=-\frac{%
\epsilon }{2}X^{1i} \\ 
\\ 
\delta X^{2i}=\frac{\epsilon }{2}X^{2i}\qquad ,\qquad \delta X^{3i}=\frac{%
\epsilon }{2}X^{3i}%
\end{array}
\label{e12}
\end{equation}%
and subsequently%
\begin{equation}
\begin{array}{l}
\delta z^{0}=\epsilon z^{0}\qquad ,\qquad \delta z^{1}=0 \\ 
\\ 
\delta z^{\widetilde{0}}=\epsilon z^{\widetilde{0}}\qquad ,\qquad \delta z^{%
\widetilde{1}}=0%
\end{array}
\label{e13}
\end{equation}%
We find that the condition of scale invariance makes the static LCR-manifold
completely trivial with $a=0$ and $f(r)=const$. This indicates that the
static LCR-manifold cannot be a state in a Hilbert space with non-broken
scale invariance.

We will now consider the CASE II (degenerate) algebraic surface $%
Z^{0}Z^{1}=0 $ which implies the conditions $X^{11}=0=X^{02}$. In this case
the convenient structure coordinates are%
\begin{equation}
\begin{array}{l}
z^{0}=i\frac{X^{21}}{X^{01}}\quad ,\quad z^{1}=\frac{X^{31}}{X^{01}}\quad
,\quad z^{\widetilde{0}}=i\frac{X^{32}}{X^{12}}\quad ,\quad z^{\widetilde{1}%
}=\frac{X^{22}}{X^{12}} \\ 
\end{array}
\label{e14}
\end{equation}%
which transform as follows%
\begin{equation}
\begin{array}{l}
\delta z^{0}=\epsilon ^{0}\qquad ,\qquad \delta z^{1}=0\qquad ,\qquad \delta
z^{\widetilde{0}}=\epsilon ^{0}\qquad ,\qquad \delta z^{\widetilde{1}}=0 \\ 
\\ 
\delta z^{0}=0\qquad ,\qquad \delta z^{1}=i\epsilon ^{12}z^{1}\qquad ,\qquad
\delta z^{\widetilde{0}}=0\qquad ,\qquad \delta z^{\widetilde{1}}=-i\epsilon
^{12}z^{\widetilde{1}}%
\end{array}
\label{e15}
\end{equation}%
under time translation and z-rotation. I found the following invariant
LCR-surface 
\begin{equation}
\begin{array}{l}
z^{0}=u=t-r\qquad ,\qquad z^{\widetilde{0}}=v=t+r \\ 
\\ 
z^{\widetilde{1}}=\overline{f(r)}\ \overline{z^{1}} \\ 
\end{array}
\label{e16}
\end{equation}%
where $f(r)$\ is an arbitrary complex function. One may easily compute the
corresponding tetrad up to their arbitrary factors $N_{1}$, $N_{2}$ and $%
N_{3}$.%
\begin{equation}
\begin{array}{l}
\ell =N_{1}du \\ 
\\ 
n=N_{2}dv \\ 
\\ 
m=N_{3}[-z^{1}\frac{df}{dr}\ dv+2fdz^{1}] \\ 
\end{array}
\label{e17}
\end{equation}%
Computing the differential forms of this tetrad we find that the $\ell $ and 
$n$ relative invariants vanish. Hence these LCR-structures cannot be
equivalent to those of the CASE I.

The dilation transformation of the new structure coordinates is%
\begin{equation}
\begin{array}{l}
\delta z^{0}=\epsilon z^{0}\qquad ,\qquad \delta z^{1}=\epsilon z^{1} \\ 
\\ 
\delta z^{\widetilde{0}}=\epsilon z^{\widetilde{0}}\qquad ,\qquad \delta z^{%
\widetilde{1}}=\epsilon z^{\widetilde{1}}%
\end{array}
\label{e18}
\end{equation}%
We find that the condition of scale invariance makes this LCR-manifold
completely trivial too with $f(r)=const$. This indicates that the static
LCR-manifold cannot be a state in a Hilbert space with non-broken scale
invariance.

We will now consider stationary configurations $X^{mi}$ which move with the
light velocity. Then $X^{mi}$ satisfy the relations%
\begin{equation}
\delta X^{mi}=i\frac{\epsilon }{2}[P_{0}+P_{3}]_{n}^{m}X^{ni}  \label{e19}
\end{equation}%
which imply%
\begin{equation}
\begin{array}{l}
\delta X^{0i}=0\qquad ,\qquad \delta X^{1i}=0 \\ 
\\ 
\delta X^{2i}=0\qquad ,\qquad \delta X^{3i}=-i\epsilon X^{1i}%
\end{array}
\label{e20}
\end{equation}%
In this case the most general quadratic polynomial, which is invariant under
the above transformations is%
\begin{equation}
(bZ^{0}+Z^{2})Z^{1}=0  \label{e21}
\end{equation}

The two solutions are $X^{11}=0$ and $X^{22}=-bX^{02}$. In this case we
cannot use the (\ref{e14}) definitions of the structure coordinates. Instead
we may use the following structure coordinates%
\begin{equation}
\begin{array}{l}
z^{0}=i\frac{X^{21}}{X^{01}}\quad ,\quad z^{1}=-i\frac{X^{31}}{X^{01}}\quad
,\quad z^{\widetilde{0}}=i\frac{X^{32}}{X^{12}}\quad ,\quad z^{\widetilde{1}%
}=\frac{X^{02}}{X^{12}} \\ 
\end{array}
\label{e22}
\end{equation}%
Then they transform as follows under the above (\ref{e20}) and the
z-rotation (\ref{e4}) transformations%
\begin{equation}
\begin{array}{l}
\delta z^{0}=0\qquad ,\qquad \delta z^{1}=0\qquad ,\qquad \delta z^{%
\widetilde{0}}=\epsilon \qquad ,\qquad \delta z^{\widetilde{1}}=0 \\ 
\\ 
\delta z^{0}=0\qquad ,\qquad \delta z^{1}=i\epsilon ^{12}z^{1}\qquad ,\qquad
\delta z^{\widetilde{0}}=0\qquad ,\qquad \delta z^{\widetilde{1}}=-i\epsilon
^{12}z^{\widetilde{1}}%
\end{array}
\label{e23}
\end{equation}

The asymptotic flatness conditions imply the following LCR-manifold%
\begin{equation}
\begin{array}{l}
U=0\quad ,\quad V=bz^{\widetilde{1}}\overline{z^{\widetilde{1}}} \\ 
\\ 
z^{\widetilde{1}}=\overline{f(u)}\ \overline{z^{1}} \\ 
\end{array}
\label{e24}
\end{equation}%
with $b$ a real constant and $f(u)$ an arbitrary complex function. One may
easily compute the corresponding tetrad up to their arbitrary factors $N_{1}$%
, $N_{2}$ and $N_{3}$.%
\begin{equation}
\begin{array}{l}
\ell =N_{1}du \\ 
\\ 
n=N_{2}[dv+ib(f^{\prime }\overline{f}-\overline{f^{\prime }}f)z^{1}\overline{%
z^{1}}du+ibf\overline{f}(\overline{z^{1}}dz^{1}-z^{1}d\overline{z^{1}}) \\ 
\\ 
m=N_{3}d(fz^{1}) \\ 
\end{array}
\label{e25}
\end{equation}%
Computing the differential forms of this tetrad we find that the relative
invariants, which correspond to $\ell $ and $m$ forms, vanish.

The dilation transformation of the new structure coordinates is%
\begin{equation}
\begin{array}{l}
\delta z^{0}=\epsilon z^{0}\qquad ,\qquad \delta z^{1}=\epsilon z^{1} \\ 
\\ 
\delta z^{\widetilde{0}}=\epsilon z^{\widetilde{0}}\qquad ,\qquad \delta z^{%
\widetilde{1}}=0%
\end{array}
\label{e26}
\end{equation}%
We find that the condition of scale invariance makes this LCR-manifold
completely trivial too (with $b=0$). This indicates that this stationary
LCR-manifold cannot be a state in a Hilbert space with non-broken scale
invariance too.

In section III, we saw that regularity of the compatible class of metrics
restricts the number of sheets up to four. Hence regular gravitational
content is expected to be derived from up to four degree Kerr polynomials.
Looking for static axially symmetric four degree polynomials, I found only
some degenerate ones. These cases seem to exhaust the static and stationary
axially symmetric LCR-manifolds, which could correspond to "stable
particles" at the quantum level.

Apparently the next step was to look for non-static axially symmetric
LCR-manifolds, which could correspond to "unstable particles". This turned
out to be too complicated.

\section{ON GEOMETRIC\ QUANTIZATION\ OF\ LCR-MANIFOLDS}

\setcounter{equation}{0}

The 4-dimensional LCR-manifolds are totally real submanifolds of an
8(real)-dimensional complex manifold. In the present section I will describe
a class of Kaehler metrics which is reduced to the class of lorentzian
metrics in the manifold. Besides, the LCR-manifolds are lagrangian
submanifolds to the corresponding symplectic manifold. These two properties
suggest to consider the ambient complex manifold as the phase space of the
LCR-manifold and look for possible applications of the geometric quantization%
\cite{WOOD}.

It is well known that the phase space of a free particle is the cotangent
bundle of a 4-dimensional space with the symplectic form

\begin{equation}
\begin{array}{l}
\omega =dp_{a}\wedge dq^{a}=d(p_{a}dq^{a}) \\ 
\end{array}
\label{q1}
\end{equation}%
Considering the vertical polarization with leaves $q^{a}=const$ , the
geometric quantization generates a Hilbert space of square-integrable
functions $\psi (q)$, where the operators are

\begin{equation}
\begin{array}{l}
\widehat{q}^{a}=q^{a}\quad ,\quad \widehat{p}_{a}=i\hbar \frac{\partial }{%
\partial q^{a}} \\ 
\end{array}
\label{q2}
\end{equation}%
The hamiltonian $h=\eta ^{ab}p_{a}p_{b}$ describes the submanifolds which
preserve the Poincar\'{e} representations with a given mass. These are the
particle trajectories in the phase space. The eigenvector equation of the
quantized hamiltonian $\widehat{h}$ is the Klein-Gordon equation. The
corresponding quantum scalar field describes the Poincar\'{e} representation
in the precise Hilbert space.

We may define the complex coordinates $r^{a}=q^{a}+i\eta ^{ab}p_{b}$.\ Then
the symplectic form becomes

\begin{equation}
\begin{array}{l}
\omega =\frac{-i}{2}\eta _{ab}dr^{a}\wedge d\overline{r^{b}}=i\frac{\partial
^{2}S}{\partial r^{a}\partial \overline{r^{b}}}dr^{a}\wedge d\overline{r^{b}}
\\ 
\\ 
S=\frac{(\overline{r^{a}}-r^{a})(\overline{r^{b}}-r^{b})\eta _{ab}}{4} \\ 
\end{array}
\label{q3}
\end{equation}%
The phase space is a Kaehler manifold and the corresponding complex
polarization is not positive definite.

In the ambient complex manifold of a general LCR-structure (\ref{i9}), I
consider the following Kaehler metric and corresponding symplectic form \ 
\begin{equation}
\begin{array}{l}
ds^{2}=\frac{\partial ^{2}S}{\partial z^{a}\partial \overline{z^{b}}}dz^{a}d%
\overline{z^{b}}\quad ,\quad \omega =i\frac{\partial ^{2}S}{\partial
z^{a}\partial \overline{z^{b}}}dz^{a}\wedge d\overline{z^{b}} \\ 
\\ 
S=A^{2}\rho _{11}\rho _{22}-B^{2}\rho _{12}\overline{\rho _{12}} \\ 
\end{array}
\label{q4}
\end{equation}%
where $A$\ and $B$\ are arbitrary non vanishing functions of the structure
coordinates. Using the derivation relations (\ref{i11}) of the tetrad of the
LCR-structure, we find that the induced metric belongs to the class of
compatible metrics. Besides the symplectic form $\omega $ vanishes on the
LCR-submanifold. Hence this submanifold is lagrangian relative to this class
of symplectic forms. In fact I considered the precise Kaehler metric in
order the LCR-manifold to become lagrangian submanifold of the ambient
complex manifold. My ultimate goal is to apply geometric quantization with a
polarization induced by the LCR-manifold. But first we must find a way to
fix the Kaehler potential $S$.

The osculation (\ref{d6}) of the LCR-manifolds imply that the Kerr functions
provide a holomorphic expression of the CR structure coordinates $%
z^{a}=f^{a}(r^{b})$ relative to the projective coordinates of $G_{2,2}$.
These holomorphic transformations do not preserve the lorentzian character
of the CR-structure, but they reveal the Poincar\'{e} transformations and
they permit us to apply the unique canonical form $y^{a}=$ $h^{a}(x)$ of the
CR-manifold. Therefore we can make the holomorphic transformation from the
structure coordinates $z^{a}$ to the $G_{2,2}$ grassmannian projective
coordinates $r^{a}$. Taking into account that the defining conditions of the
LCR-manifold can always be locally written\cite{BAOU} as $\rho
_{ij}=e_{a}^{(ij)}(y^{a}-h^{a})$ the Kaehler potential takes the form \ 
\begin{equation}
\begin{array}{l}
S=\left( y^{a}-h^{a}(x)\right) \left( y^{b}-h^{b}(x)\right) \gamma _{ab} \\ 
\end{array}
\label{q5}
\end{equation}%
where $\gamma _{ab}$ contains the arbitrary functions $A(x,y)$ and $B(x,y)$.
Notice that in the vanishing gravity approximation $h^{a}(x)\simeq 0$, the
assumption $\gamma _{ab}=\eta _{ab}$ gives the particle phase space. But in
the general case the formalism becomes too complicated. The natural
foliation $x^{a}=const$ of the ambient complex manifold is not generally a
polarization. On the other hand, the complex polarization is not generally
quantizable because the scalar product \ 
\begin{equation}
\begin{array}{l}
(\Phi ,\Psi )=\int \overline{\Phi }\Psi e^{-S}\omega ^{n} \\ 
\end{array}
\label{q6}
\end{equation}%
is not generally an integrable quantity ($S$ is not positive). A solution to
this problem could be the possibility to holomorphically transfer the
LCR-manifolds inside the $SU(2,2)$ classical domain, which would permit us
to restrict the integration to a finite region of $%
\mathbb{C}
^{4}$. But I have not yet found a way to choose a physically interesting
regular Kaehler potential $S$. Concluding this section, I want to point out
that the flat spacetime case (\ref{q3}) indicates that $S$ should be related
to the Bergman kernel of the domain bounded by the LCR-manifold.

\section{TOWARDS A LCR STRUCTURE\ QUANTUM\ FIELD\ THEORY}

\setcounter{equation}{0}

The path integral quantization gives the state $\Psi (x^{\prime \prime
},t^{\prime \prime })$ at time $t^{\prime \prime }$\ 
\begin{equation}
\begin{array}{l}
\Psi (x^{\prime \prime },t^{\prime \prime })=e^{-\frac{i}{\hslash }%
(t^{\prime \prime }-t^{\prime })H}\Psi (x^{\prime \prime },t^{\prime })=\int
dx^{\prime }\ K(x^{\prime \prime },t^{\prime \prime };x^{\prime },t^{\prime
})\ \Psi (x^{\prime },t^{\prime }) \\ 
\\ 
K(x^{\prime \prime },t^{\prime \prime };x^{\prime },t^{\prime })=\overset{%
x(t^{\prime \prime })=x^{\prime \prime }}{\underset{x(t^{\prime })=x^{\prime
}}{\int }}[dx(t)]\ e^{iI[x(t)]} \\ 
\end{array}
\label{p1}
\end{equation}%
as the functional integral over all the trajectories in the time-interval $%
(t^{\prime },t^{\prime \prime })$. In the case of a 2-dimensional generally
covariant QFT (like the Polyakov string action) the integration is performed%
\cite{POL} over the fields $X^{\mu }(x,t)$ and the 2-dimensional surfaces $M$
with "initial" $B_{i}$ and "final" $B_{f}$ boundaries.\ 
\begin{equation}
\begin{array}{l}
\Psi (B_{f},t^{\prime \prime })=e^{-\frac{i}{\hslash }(t^{\prime \prime
}-t^{\prime })H}\Psi (B_{f},t^{\prime })=\underset{topol}{\sum }\int
[dB_{i}]\ K(B_{f},t^{\prime \prime };B_{i},t^{\prime })\ \Psi
(B_{i},t^{\prime }) \\ 
\\ 
K(B_{f},t^{\prime \prime };B_{i},t^{\prime })=\underset{\partial M=B_{i}\cup
B_{f}}{\int }\![dM][d\phi (x)]\ e^{iI[\phi (x)]} \\ 
\end{array}
\label{p2}
\end{equation}%
where we sum over the topologies of the 2-dimensional surfaces. General
2-dimensional covariance implies $H=0$. The recent interest to String Field
Theory gave a strong push to the systematic study of functional integrals
over (2-dimentional) Riemann surfaces. Integration over simple surfaces with
one boundary cycle provide the vacuum and the string states in the
Schrodinger representation. Integration over complex structures on the
simple cylinder gives the first approximation to the string propagator,
while the addition of handles provide the "loop-diagrams". The splitting of
a cylinder-like 2-dimensional surface into two cylinder-like surfaces ("pair
of pants" string diagram) is viewed as the fundamental coupling constant in
String Quantum Field Theory. The corresponding splitting of the string modes
is considered as the emission of two new strings by the original string.

Using the properties of the functional integrations (path integrals), Segal
created an axiomatic formulation of 2-dimensional Conformal Field Theory.
Following analogous steps, Atiyah created an axiomatic formulation of the
Topological Quantum Field Theories. The purpose of this section is to show
that analogous properties exist in the present model, which could lead to a
LCR-QFT description of the LCR-manifolds

In the present case, we have the following kernel%
\begin{equation}
\begin{array}{l}
K(B_{f},t^{\prime \prime };B_{i},t^{\prime })=\underset{topol}{\sum }\ 
\underset{rel.inv.}{\sum }\ \underset{\partial S=B_{i}\cup B_{f}}{\int }%
[d\mu (LCR)][dA_{j\nu }]\ e^{iI_{G}[e_{\ \nu }^{a},\ A_{j\nu }]} \\ 
\end{array}
\label{p3}
\end{equation}%
where the functional integrations $[d\mu (LCR)]$ and $[dA_{j\nu }]$\ are
performed over the LCR-manifolds with initial and final boundaries, and the
gauge field connections respectively. The measure of the LCR-stuctures
moduli space is found using the integrability conditions on the tetrad and
the Faddeev-Popov technique applied to the remaining local symmetries
(diffeomorphism and tetrad-Weyl transformation). The summation includes now
the LCR-structure relative invariants in addition to the topologies, which
are not covered by the functional integration. General 4-dimensional
covariance implies $H=0$.

In the case of the String Field Theory, the functional integral of the
Polyakov action over the annulus surfaces (the cylinder-like 2-dimensional
surfaces with two circles $X_{i}^{\mu }(\sigma )$ and $X_{f}^{\mu }(\sigma )$%
\ as boundaries) provide the first approximation of the string propagator.
In the present case the functional integration over the LCR-manifolds with
two 3-dimensional boundaries $B_{i}$ and $B_{f}$\ , provide the possible
propagators with prescribed initial and final surfaces with given tetrads $%
e_{\mu }^{(i)a}$ , $e_{\mu }^{(f)a}$ and connections $A_{j\mu }^{(i)}$ , $%
A_{j\mu }^{(f)}$. The present model is much richer than the string model.
Besides the gauge field modes, which replace the string modes, there are the
initial and final CR-structures. On the other hand the sewing of
LCR-manifolds seems to be very complicated in distinction to the "simple"
sewing methods of the Riemann surfaces. I do not actually know any
systematic mathematical construction/classification of LCR-manifolds from
simple "building blocks", like the cylinder in the Riemann surfaces.
Therefore we need some "physical intuition" as starting point.

In order to get a physical picture, recall the Einstein-Infeld-Hofman (EIH)
approximation method for the derivation of the equations of motion of two
bodies. A body is considered as the gravitational field concentrated inside
a 4-dimensional "world tube". In the proper EIH approximation, this is
approximated with a singularity, while the Fock-Papapetrou approximation
considers regular configurations. The interacting two bodies are viewed as
approaching "world tubes", where the Bianchi identity is applied with an
appropriate definition of the body mass. The essential difference is that
here we consider as independent physical variable the (regular)
LCR-structure and not the metric of the manifold. That is, we consider the
class of metrics [$g_{\mu \nu }$] determined by the LCR-structure and not
each distinct metric. The splitting of a "world tube" into two "world
tubes", describes the fundamental interactions of the model.

The 3-dimensional CR-structures of $B_{i}$ and $B_{f}$ restrict the set of
integrated LCR-manifolds $M$ to those which admit a "collar" neighborhood $%
B_{f}\times \lbrack 0,1]$. This 4-dimensional "collar" is necessary to
achieve the appropriate sewing of LCR-manifolds in the definition of the
scalar product in the corresponding Hilbert space. Hence we need a
"time"-variable, which will be defined below.

We already know that a realizable LCR-manifold can be embedded into the $%
SU(2,2)$ classical domain, where (in its unbounded realization) we can
define a special coordinate system $x^{a}=\func{Re}(r^{a})$, which properly
transforms under the Poincar\'{e} group. Hence we have a time variable to
describe the evolution of the embedded 4-dimensional LCR-manifold into the
Siegel domain.

The simplest partition function is given by the integration on
LCR-structures over the closed $U(2)$ manifold. The corresponding Siegel
"real axis" covers the half of the $U(2)$ manifold. One can easily see it,
using the following transformation between the bounded Cartan domain and the
unbounded Siegel "upper half-plain" domain

\begin{equation}
\begin{array}{l}
r=i(\mathbf{1}+z)(\mathbf{1}-z)^{-1}=i(\mathbf{1}-z)^{-1}(\mathbf{1}+z) \\ 
\\ 
z=(r-i\mathbf{1})(r+i\mathbf{1})^{-1}=(r+i\mathbf{1})^{-1}(r-i\mathbf{1})%
\end{array}
\label{p4}
\end{equation}%
On the boundaries, it takes the form

\begin{equation}
\begin{array}{l}
t=\frac{\sin \tau }{\cos \tau \ -\ \cos \rho } \\ 
\\ 
x+iy=\frac{\sin \rho }{\cos \tau \ -\cos \rho }\sin \theta \ e^{i\varphi }
\\ 
\\ 
z=\frac{\sin \rho }{\cos \tau \ -\ \cos \rho }\cos \theta 
\end{array}
\label{p5}
\end{equation}%
where $t,x,y,z$ are the $R^{4}$ coordinates and $\tau ,\rho ,\theta ,\varphi 
$ are the $U(2)$ coordinates 
\begin{equation}
\begin{array}{l}
U=e^{i\tau }\left( 
\begin{array}{cc}
\cos \rho +i\sin \rho \cos \theta  & i\sin \rho \sin \theta \ e^{-i\varphi }
\\ 
i\sin \rho \sin \theta \ e^{i\varphi } & \cos \rho -i\sin \rho \cos \theta 
\end{array}%
\right) = \\ 
\\ 
\qquad =\frac{1+r^{2}-t^{2}+2it}{[1+2(t^{2}+r^{2})+(t^{2}-r^{2})^{2}]}\left( 
\begin{array}{cc}
1+t^{2}-r^{2}-2iz & -2i(x-iy) \\ 
-2i(x+iy) & 1+t^{2}-r^{2}+2iz%
\end{array}%
\right) 
\end{array}
\label{p6}
\end{equation}%
Notice that the limit $t\rightarrow \pm \infty $ with constant $x,y,z$ is
the point $U=\mathbf{1}$. Hence a state-operator correspondence%
\begin{equation}
\begin{array}{l}
\Psi _{\Phi }(B_{f},t)=\underset{topol}{\sum }\ \underset{rel.inv.}{\sum }\ 
\underset{\partial M=B_{f}}{\int }[d\mu (LCR)][dA_{j\nu }]\ e^{iI[e_{\ \nu
}^{a},\ A_{j\nu }]}\Phi (-\infty ) \\ 
\end{array}
\label{p7}
\end{equation}%
seems to exist in this model, like in the 2-dimensional conformal models.
The limit $r\rightarrow \infty $ with constant $t,\theta ,\varphi $ is the
point $U=-\mathbf{1}$. The limit $t\rightarrow \infty $ with constant
structure coordinates $z^{\alpha },\ \alpha =0,1$ is the 3-dimensional
CR-manifold (the Penrose scri+ boundary) and the limit $t\rightarrow -\infty 
$ with constant structure coordinates $z^{\widetilde{\alpha }},\ \alpha =0,1$%
\ is another 3-dimensional CR-manifold (the Penrose scri- boundary)\cite%
{P-R1984}. If these two boundaries are identified, the LCR-structures are
restricted to the periodic ones. But there are LCR-stuctures, like the
massive Kerr-Schild one, which are not periodic. Therefore these
LCR-manifolds with two boundaries seem to be the intuitive building blocks
of the interaction picture in the present cosmodynamic model. The initial
and final "collar" neighborhoods should be static axially symmetric
LCR-manifolds. I have not yet been able to compute the partition function in
any manifold.

\end{document}